\documentclass[useAMS,usenatbib]{mn2e}

\usepackage{xspace}
\usepackage{aas_macros}
\usepackage{graphicx}
\usepackage{color,url}
\usepackage{amsmath}

\def\deg{\ensuremath{^\circ}\xspace}
\def\min{\rm{'}\xspace}
\def\cep{$\beta$ Cephei\xspace}

\def\A{\AA\xspace}
\def\kms{\ensuremath{{\rm km\, s}^{-1}}\xspace}
\def\K{\ensuremath{{\rm K}}\xspace}
\def\l{\ensuremath{\ell}\xspace}
\def\m{\ensuremath{m}\xspace}

\def\SiIII{Si\,{\sc iii}\xspace}
\def\veq{\ensuremath{v_{\rm eq}}\xspace}
\def\vsini{\ensuremath{v \sin i}\xspace}
\def\cd{\ensuremath{{\rm d}^{-1}}\xspace}
\def\fo{\ensuremath{f_1}\xspace}
\def\ft{\ensuremath{f_2}\xspace}
\def\fth{\ensuremath{f_3}\xspace}
\def\ff{\ensuremath{f_4}\xspace}
\def\mHz{\ensuremath{\mu{\rm Hz}}\xspace}

\title[An asteroseismic study of the $\beta$ Cephei star 12 Lacertae]{
An asteroseismic study of the $\beta$ Cephei star 12 Lacertae:
multisite spectroscopic observations, mode identification and seismic modelling}
\author[M. Desmet et al.]
{M. Desmet$^1$\thanks{E-mail:maarten.desmet@ster.kuleuven.be},
M. Briquet$^1$\thanks{Postdoctoral Fellow of the Fund for Scientific Research, Flanders}, 
A. Thoul${^2}$\thanks{Chercheur Qualifi\'e au Fonds National de la Recherche Scientifique, Belgium.},
W. Zima$^{1}$, 
P. De Cat$^{3}$, 
G. Handler$^4$,
\newauthor
I. Ilyin$^5$, 
E. Kambe$^6$,
J. Krzesinski$^7$,
H. Lehmann$^8$, 
S. Masuda$^9$, 
P. Mathias$^{10}$, 
\newauthor
D. E. Mkrtichian$^{11}$, 
J. Telting$^{12}$,
K. Uytterhoeven$^{13}$, S. L. S.
Yang$^{14}$ 
\newauthor
and C. Aerts$^{1,15}$
\\
$^1$ Institute of Astronomy - KULeuven,
Celestijnenlaan 200D, 3001 Leuven, Belgium\\ 
$^2$ Institut d'Astrophysique et de G\'eophysique
de l'Universit\'e de Li\`ege,  17, All\'ee du 6 Ao\^ut,
4000 Li\`ege, Belgium\\
$^3$ Koninklijke Sterrenwacht van Belgi\"e, Ringlaan 3, 1180 Brussel,
Belgium\\
$^4$ Institut f\"ur Astronomie, Universist\"at Wien, T\"urkenschanzstrasse 17,
1180 Wien, Austria \\
$^5$ Astrophysical Institute Potsdam, An der Sternwarte 16
D-14482 Potsdam, Germany\\
$^6$ Okayama Astrophysical Observatory, National Astronomical
Observatory, Kamogata, Okayama 719-0232, Japan\\
$^7$ Mt. Suhora Observatory, Cracow Pedagogical University, Ul. Podchorazych
2, 30-084 Cracow, Poland\\
$^8$ Karl-Schwarzschild-Observatorium, Th\"uringer Landessternwarte, 7778
Tautenburg, Germany\\
$^9$ Tokushima Science Museum, Asutamu Land Tokushima, 45 - 22 Kibigadani,
Nato, Itano-cho, Itano-gun, Tokushima 779-0111, Japan\\
$^{10}$ UNS, CNRS, OCA, Campus Valrose, UMR 6525 H. Fizeau, F-06108 Nice Cedex 2,
France\\
$^{11}$ Astronomical Observatory of Odessa National University, Marazlievskaya,
1v, 65014 Odessa, Ukraine\\
$^{12}$ Nordic Optical Telescope, Apartado 474, 38700 Santa Cruz de La Palma,
Spain\\
$^{13}$ Instituto de Astrof\'{\i}sica de Canarias, Calle Via L\'actea s/n,
38205 La Laguna (TF, Spain)\\
$^{14}$ Department of Physics and Astronomy, Universtity of Victoria,
Victoria, BC V8W 3P6, Canada\\
$^{15}$ Department of Astrophysics, University of Nijmegen, PO Box
9010, 6500 GL Nijmegen, The Netherlands
}
\begin{document}

\date{Accepted 2008. Received 2008}

\pagerange{\pageref{firstpage}--\pageref{lastpage}} \pubyear{2002}

\maketitle

\label{firstpage}


\begin{abstract}
We present the results of a spectroscopic multisite campaign for the $\beta$
Cephei star 12 (DD) Lacertae. Our study is based on more than thousand
high-resolution high S/N spectra gathered with 8 different telescopes in a
time span of 11 months. In addition we make use of numerous archival
spectroscopic measurements. We confirm 10 independent frequencies recently
discovered from photometry, as well as harmonics and combination
frequencies. In particular, the SPB-like g-mode with frequency 0.3428\,\cd
reported before is detected in our spectroscopy. We identify the four main
modes as $(\l_1,\m_1)=(1,1)$, $(\l_2,\m_2)=(0,0)$, $(\l_3,\m_3)=(1,0)$ and
$(\l_4,\m_4)=(2,1)$ for $f_1 = 5.178964$\,\cd, $f_2 = 5.334224$\,\cd, $f_3 =
5.066316$\,\cd and $f_4 =  5.490133$\,\cd, respectively.  Our seismic
modelling shows that \ft is  likely
the radial first overtone and that the core
overshooting parameter $\alpha_{\rm{ov}}$ is lower than 0.4 local pressure
scale heights. 

\end{abstract}


\section{Introduction}
12 (DD) Lacertae (12 Lac) is one of the best observed $\beta$ Cephei stars, a
class of variable early B-type stars whose light and radial velocity changes
are due to gravity and pressure-mode pulsations of low radial order
\citep{2005yCat..21580193S}. The variability of 12 Lac has been known for one
century during which the star has been extensively studied. We refer to
\citet{2006MNRAS.365..327H} for a very detailed summary of the work
accomplished in the past. 

Despite the numerous earlier studies, safe mode identification of the observed
modes was not achieved until recently. Thanks to an intensive photometric
multisite campaign, \citet{2006MNRAS.365..327H} unambiguously identified the
$\ell$-values of the five modes with the highest photometric amplitudes. In
addition, they found constraints on $\ell$ for the six other independent modes
detected in their dataset. Their $\ell$-identifications ruled out the
assumption previously adopted for stellar modelling
\citep[e.g.][]{1999A&A...341..480D} that three of the strongest modes, almost
equidistant, belong to the same multiplet. Indeed, these three modes are
actually associated to three different values of the degree $\ell$. 

Obviously, reliable empirical mode identification was indispensable before any
attempt of in-depth seismic modelling of 12 Lac. To complement the photometric
results, a spectroscopic multisite campaign has also been devoted to the star.
The additional constraints are presented in this paper. They mainly concern
the identification of $m$-values for the strongest modes and the derivation of
the stellar equatorial velocity. A detailed abundance analysis of 12 Lac was
already presented in \citet{2006A&A...457..651M}, showing that the abundances
of all considered chemical elements are indistinguishable from the values
found for OB dwarfs in the solar neighbourhood.

Besides our line-profile study, we also describe a detailed stellar modelling
based on all available observational results, state-of-the-art numerical tools
and up-to-date physical inputs appropriate to model $\beta$ Cephei stars, as
explained in detail in \citet{Miglio}. \citet{2008MNRAS.385.2061D} already
computed models for 12 Lac, making some assumptions and restrictions not
supported by our study. In the present paper, we discuss our conclusions which
differ from those of \citet{2008MNRAS.385.2061D}.


\begin{figure}
\rotatebox{-90}{\resizebox{!}{\columnwidth}{\includegraphics{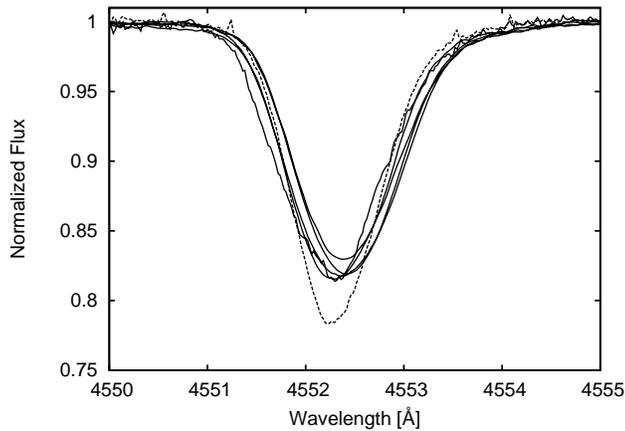}}}
\caption{The average profiles of the \SiIII $\lambda 4553$ \A line from each
observatory.  The dashed line shows the average profile from the BAO
observatory. This profile deviates from all the others
because the relatively small number of spectra implies that the beat cycle is
not well covered (see
Table\,\ref{TAB:observatories}).
}
\label{FIG:observatories_average}
\end{figure}
\begin{figure}
\begin{center}
\rotatebox{0}{\resizebox{!}{\columnwidth}{\includegraphics{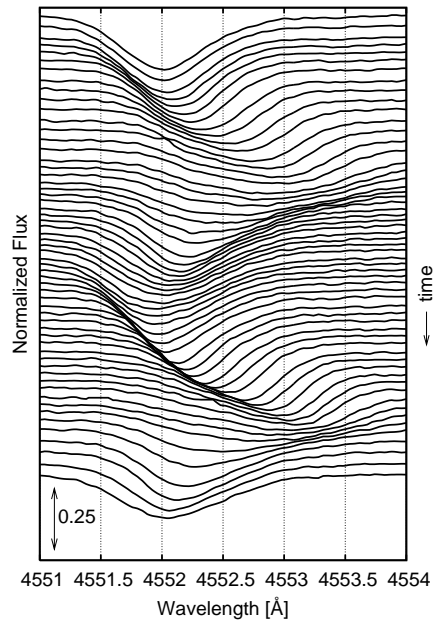}}}
\caption{The \SiIII line profiles of 1 night (JD 2452898)  
taken with the McDonald telescope.
The arrow indicates the flux scale of one spectrum. } 
\label{FIG:1night}
\end{center}
\end{figure}
\begin{figure*}
\begin{center}
\rotatebox{-90}{\resizebox{!}{\hsize}{\includegraphics{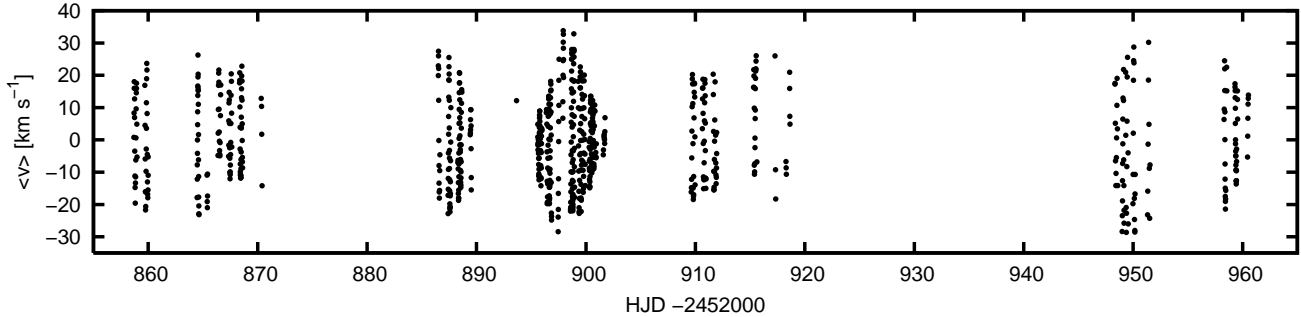}}}
\caption{The densest part of the time series of the  \SiIII radial
velocities ($\langle v^1 \rangle$) of 12 Lac derived from each spectrum taken during the dedicated
multisite campaign (7 Aug. 2003 -- 17 Nov. 2003).  } 
\label{FIG:RV} 
\end{center} 
\end{figure*}

\section{Observations and data reduction}

The data originate from a spectroscopic multisite campaign dedicated to our
studied star. In addition, we added the 903 spectra of
\citet{1994A&A...289..875M}, which allowed us to increase the time span and,
thus, to achieve a better frequency accuracy. 
In total 1820
observations were gathered using 9 different small- to medium-sized telescopes
spread over the northern continents. Table \ref{TAB:observatories} summarizes
the logbook of our spectroscopic data. 
The resolution ($\lambda/\Delta \lambda$) of the instruments ranged from
30\,000 to 80\,000 and the average S/N ratio near 4500 \A between 180 and 380.


\begin{table*}
 \centering
 \begin{minipage}{180mm}
  \caption{Log of our spectroscopic multisite campaign. The Julian Dates are
given in days, $\Delta T$ denotes the time-span expressed in days, $N$ is the
number of spectra and S/N denotes the average signal-to-noise ratio for each
observatory measured
at the continuum between 4500 and 4551 \A.}
\label{TAB:observatories}
  \begin{tabular}{@{}lrrccccccl@{}}
  \hline
  \hline
  Observatory  & Long. & Lat. & Telescope & \multicolumn{2}{c}{Julian Date}  
& \multicolumn{3}{c}{Data amount and quality} &  Observer(s) \\
(Name of the instrument; resolution) &&&& Begin & End  & $\Delta T$  &  N & S/N & \\
 \hline
 \hline
  2003-2004   &&&&\multicolumn{2}{c}{-2450000}&&&&\\
\hline
 Apache Point Observatory  & $-105\deg49\min$ & $+32\deg46\min$
&3.5-m&2927&2928&2 & 93 & 300 & JK \\
\hspace{2mm}(ARC; 31500)&&&&&&&&&\\
 Bohyunsan Astronomical Observatory  & $+128\deg58\min$ &$+36\deg09\min$
&1.9-m& 3109 & 3109 &1 &27 & 180 & DM \\
\hspace{2mm}(BOES; 30000)&&&&&&&&&\\
 Dominion Astrophysical Observatory &$-123\deg25\min$ &$+48\deg31\min$
&1.2-m&2858 & 2859& 2 &66 & 250 & SY\\
\hspace{2mm}(45000) &&&&2909&2911&3 &42&&\\
 McDonald Observatory  & $-104\deg01\min$ & $+30\deg40\min$ & 2.1-m & 2893 &
2901& 9 &210 & 380 & GH \\
\hspace{2mm}(Sandiford; 60000) &  &  & 2.7-m &3195 &3198&4 & 31 & 350  & GH \\
 Nordic Optical Telescope & $-17\deg53\min$ & $+28\deg45\min$ & 2.6-m
&2947&2951& 5 &32 & 370 & KU, JT, II\\
\hspace{2mm}(SOFIN; 80000) &  &  &  &2957 &2960 &4 &47 &  &  \\
 Okayama Astrophysical Observatory & $+133\deg35\min$ & $+34\deg34\min$ &
1.88-m & 3201 &3202& 2 & 7 & 180 & SM \\
\hspace{2mm}(HIDES; 685000) &  &  &  &3205 &3205&1 & 6 &  &  \\
 &  &  &  &3208 &3210& 3& 17 &  &  \\
 Th\"uringer Landessternwarte Tautenburg & $+11\deg42\min$ & $+50\deg58\min$ &
2-m &2863 &2869& 7& 103 & 240  & HL \\
\hspace{2mm}(67000) &  &  &  &2886 &2889&4 & 100 &  &  \\
 &  &  &  &2895 &2900& 6& 105 &  &  \\
 &  &  &  &2914 &2918& 5& 31 &  &  \\
\hline
  1990-1992   &&&&\multicolumn{2}{c}{-2440000} &&&&\\
\hline
 Observatoire de Haute-Provence & $+5\deg42\min$ & $+43\deg55\min$ & 1.52-m &
8135 & 8135& 1 &70  & 220  &PM \\
\hspace{2mm}(AURELIE; 42000) &  &  &  &8224 &8224&1 & 52 &  &  \\
 &  &  &  &8227 &8227&1 & 35 &  &  \\
 &  &  &  &8845 &8858&14 & 582 &  &  \\
 &  &  &  &8884 &8889& 6& 164 &  &  \\
\hline
\hline
 Total&  &  &  & & &82 & 1820 &  &  \\
\hline
\hline
\end{tabular}
\end{minipage}
\end{table*}

All data were subjected to the normal reduction process, which consists of
de-biasing, background subtraction, flat-fielding and wavelength calibration .
Finally, the heliocentric corrections were computed, and all spectra were
normalized to the continuum by fitting a cubic spline function.
    
For our study, we considered the \SiIII triplet around 4567 \A because its
characteristics simplify the modelling of the line-profile variations, which
we use for mode identification. Indeed, these silicon lines are sufficiently
strong without being much affected by blending. Moreover, they are dominated
by temperature broadening, such that the intrinsic profile can simply be
modelled with a gaussian. Finally, they are not too sensitive to temperature
variations \citep[see][]{2002A&A...385..572D,2003SSRv..105..453A}, so that
neglecting them remains justified.    

The similar average radial velocity of about $-15$ \kms for 1990-1992,
2003-2004 and earlier measurements \citep[e.g.][]{1953GCRV..C......0W}
definitely excludes the possibility of 12 Lac being a spectroscopic binary.
Besides the pulsation effect, the different centroids are due to the different
zero points of the different telescopes.  Before extracting the pulsation
information from the whole dataset, it is necessary to correct for this
effect. The spectra were shifted in such a way that the constant of a
least-squares sine fit using the first three dominant modes is put to the same
value for each observatory.

In Fig.\,\ref{FIG:observatories_average} we superimpose the average profiles
of the \SiIII $\lambda 4553$ \A line computed for each observatory separately,
after correction for the different zero points.  We note that the centroid of
the lines are in good agreement but not exactly at the same position.  Larger
differences can be seen for the width and the depth of the lines.  Such
deviant average profiles are due to the limited time spread of the
observations together with the multiperiodic character of the pulsations. In
the case of a similar spectroscopic multisite campaign devoted to $\nu$
Eridani, \citet{2004MNRAS.347..463A} observed similar differences, which are
not caused by instrumental and/or signal-to-noise ratio effects.

Fig.\,\ref{FIG:1night} displays the large and complex radial-velocity variations
of 12 Lac.  The plotted spectra were taken with the McDonald telescope during
only one night. In Fig.\,\ref{FIG:RV}, we show the radial velocities for the
densest part of the multisite campaign (7 Aug. 2003 -- 17 Nov. 2003). A clear beating pattern is seen.

\section{Frequency analysis}
We performed a frequency analysis on the first three velocity moments $\langle
v^1\rangle$, $\langle v^2\rangle$ and $\langle v^3\rangle$ \citep[see][for a
definition of the moments of a line profile]{1992A&A...266..294A} of the
\SiIII $\lambda 4553$ \A line by means of the program Period04
\citep{2005CoAst.146...53L}. For some \cep stars
\citep[see][]{1997A&A...322..493T,2004A&A...416.1069S,2005MNRAS.362..619B} a
two-dimensional frequency analysis on the spectral lines led to additional
frequencies compared to a one-dimensional frequency search in the moments.
For 12 Lac 1D analysis on $\langle
v^2\rangle$ and $\langle v^3\rangle$ did not reveal additional independent
periodicities compared to those present in the first moment.  
We also do not find additional frequencies in the 2D frequency analysis.
The 2D analysis is more sensitive to the detection of 
high-degree modes than a 1D analysis so here we have already an indication 
that the frequencies are low-degree modes. 

In what follows
we only describe our frequency analysis on $\langle v^1\rangle$, which is the
radial velocity from which the fitted averaged radial velocity is subtracted. 

The frequencies were determined with the standard method of prewhitening the
data. In addition we made use of a procedure available in Period04 which
allows to improve the detected frequency from non-linear least-squares fitting
with the maxima in the Fourier transform as starting values. At each step of
prewhitening we subtracted a theoretical multi-sine fit with the amplitudes,
phases and also optimized frequencies that yielded the smallest residual
variance. All peaks exceeding an amplitude signal-to-noise ratio above 4 in
the Fourier periodogram \citep{1993A&A...271..482B,1999A&A...349..225B} were
retained. The noise level was calculated from the average amplitude, computed
from 
the residuals, in a 2\,\cd interval centered on the frequency of interest.

\begin{table}
\centering
  \caption{Frequencies and radial velocity amplitudes of the first moment of the \SiIII
$\lambda 4553$\,\A line together with their S/N ratio (we refer to the text
for explanation). Error estimates \citep{1999DSSN...13...28M}  
for the independent frequencies range from $\pm
0.000002\,\cd$ for \fo to $\pm 0.00002\,\cd$ for $f_9$. The error on the amplitude is
$0.01$ \kms.
}
\label{TAB:FREQ}
  \begin{tabular}{@{}lcccc@{}}
\hline \hline
ID &\multicolumn{2}{c}{Frequency} & Amplitude in $\langle v^1\rangle$ & S/N \\
& [\cd] & [\mHz] & [\kms] &\\
\hline
\fo  &          5.178964   & 59.941713   &   14.50 & 99.6  \\
\ft  &          5.334224   & 61.738704   &    7.70 & 52.2  \\
\fth &          5.066316   & 58.637917   &    6.26 & 42.7  \\
\ff  &          5.490133   & 63.543206   &    2.61 & 17.5  \\
$f_g$    &      0.342841   & 3.968067    &    1.34 & 7.1   \\
$f_5    $ &     4.241787   & 49.09475   &    0.91 & 6.7   \\
$f_6    $ &     5.218075   & 60.394387   &    0.84 & 5.8   \\
$f_7    $ &     6.702318   & 77.573125   &    0.62 & 4.4   \\
$f_8    $ &     7.407162   & 85.731042   &    0.67 & 4.9   \\
$f_9    $ &     5.84511   & 67.65184   &    0.79   & 5.2   \\[1ex]
$2f_1   $ &     10.35814  &  119.88590 &    0.63   & 4.5   \\
$f_2+f_3$ &     10.40056  &  120.37693 &    0.72   & 3.8   \\
$f_1+f_2$ &     10.51319  &  121.68044 &    2.59   & 19.7  \\
$2f_1+f_3$&     15.42400  &  178.51860 &    0.72   & 6.6   \\
$f_1+f_2+f_3$  &15.57950  &  180.31838 &    0.71   & 6.4   \\
\hline

\end{tabular}
\end{table}

\begin{figure*}
\begin{center}
\rotatebox{-90}{\resizebox{!}{\hsize}{\includegraphics{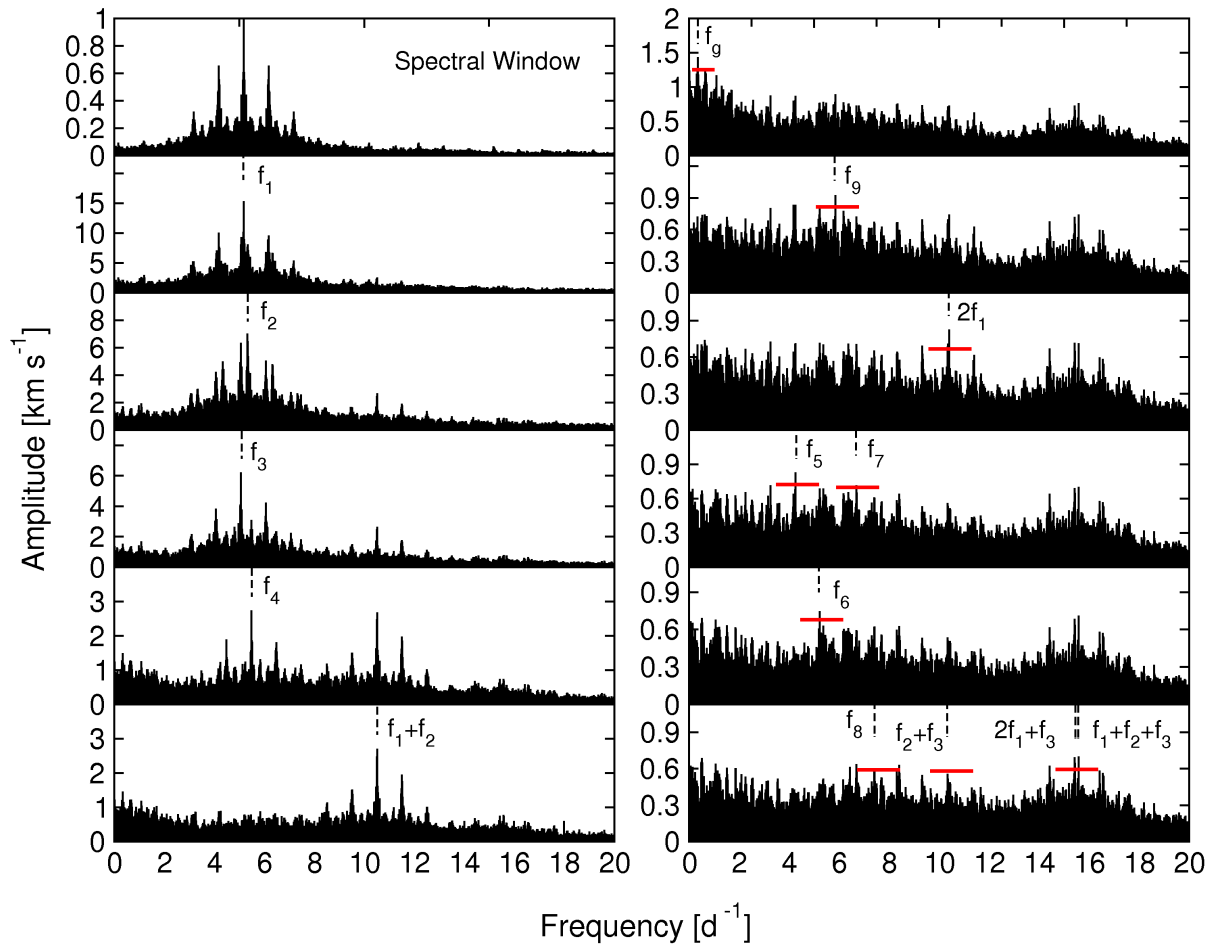}}}
\caption{ Amplitude spectra of 12 Lac computed from the first moment of the
\SiIII $\lambda
4553$ \A line. The uppermost panel shows the spectral window of the data.  All
subsequent panels show the periodograms in different stages of
prewhitening. The significance limit in red is calculated according to 4 times
the noise over a 2\,\cd-range around the significant frequency.
}
\label{FIG:AMPLITUDESPECTRA}
\end{center}
\end{figure*}

The detected frequencies are listed in Table \ref{TAB:FREQ} and the Fourier
periodograms are shown in Fig.\,\ref{FIG:AMPLITUDESPECTRA}.  Our frequency
analysis reaffirmes the presence of the already well-known 5 main frequencies
\citep{1994A&A...289..875M} together with the new independent signals and 
combination frequencies discovered by \citet{2006MNRAS.365..327H}. We note
that one additional independent frequency ($f_p = 5.30912$\,\cd) was found by
the latter authors  and not by us. 
An important result is that we clearly recover the
low-frequency signal possibly originating from a $g$-mode. Such a SPB-like
oscillation does not seem to be uncommon among $\beta$ Cephei stars. For
instance, it was also observed in 19 Mon \citep{2002MNRAS.333..952B}, in $\nu$
Eridani \citep{2004MNRAS.347..454H}, in $\delta$ Ceti
\citep{2006ApJ...642..470A} and in V1449~Aql (Briquet et  al., in
preparation).  We also point out that such hybrid SPB-$\beta$ Cep pulsators
are theoretically predicted for hotter stars than previously thought
(\citealp{2007MNRAS.375L..21M} versus \citealp{1999AcA....49..119P}).

\begin{figure}
\begin{center}
\rotatebox{-90}{\resizebox{!}{\columnwidth}{\includegraphics{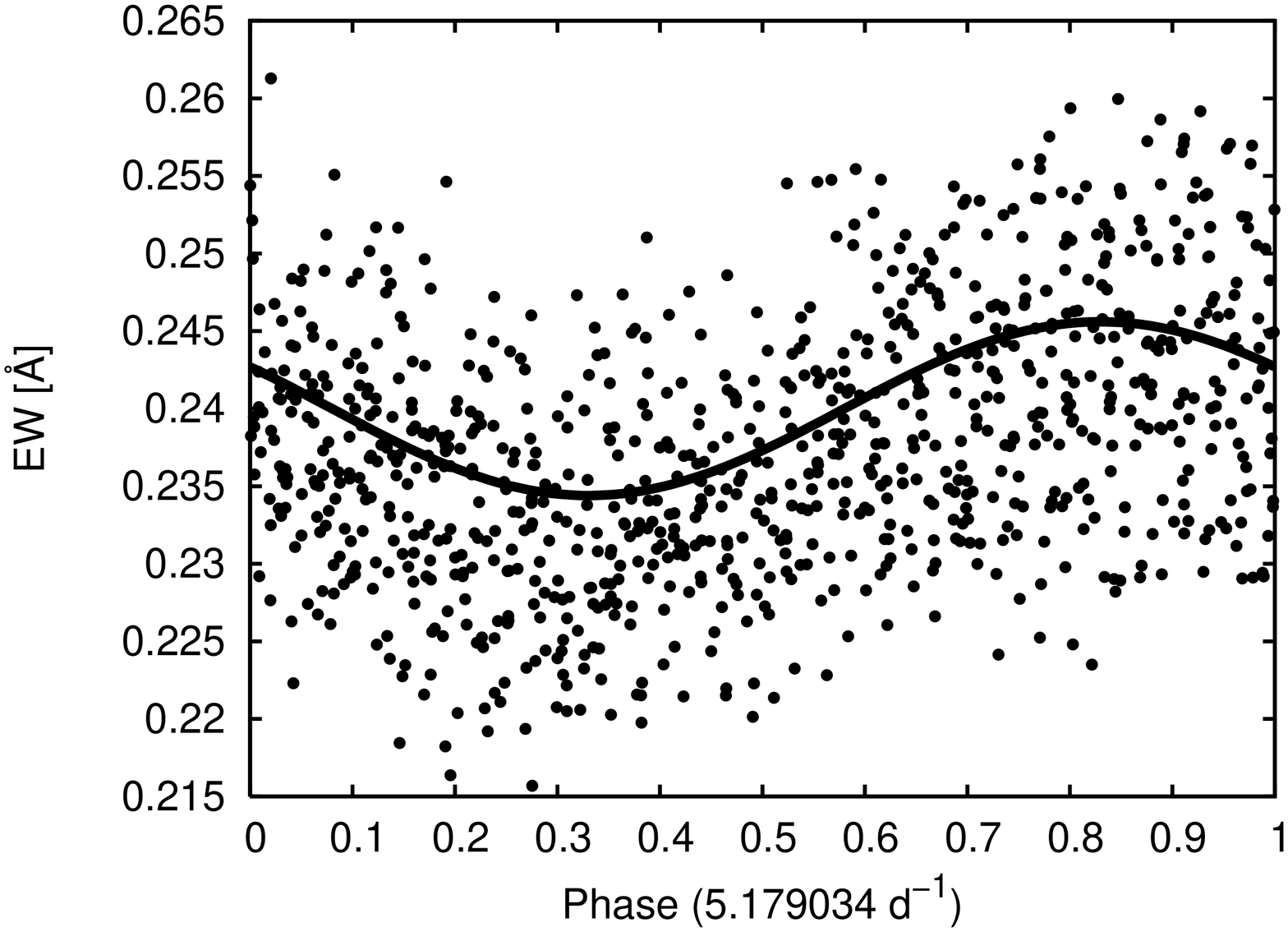}}}
\rotatebox{-90}{\resizebox{!}{\columnwidth}{\includegraphics{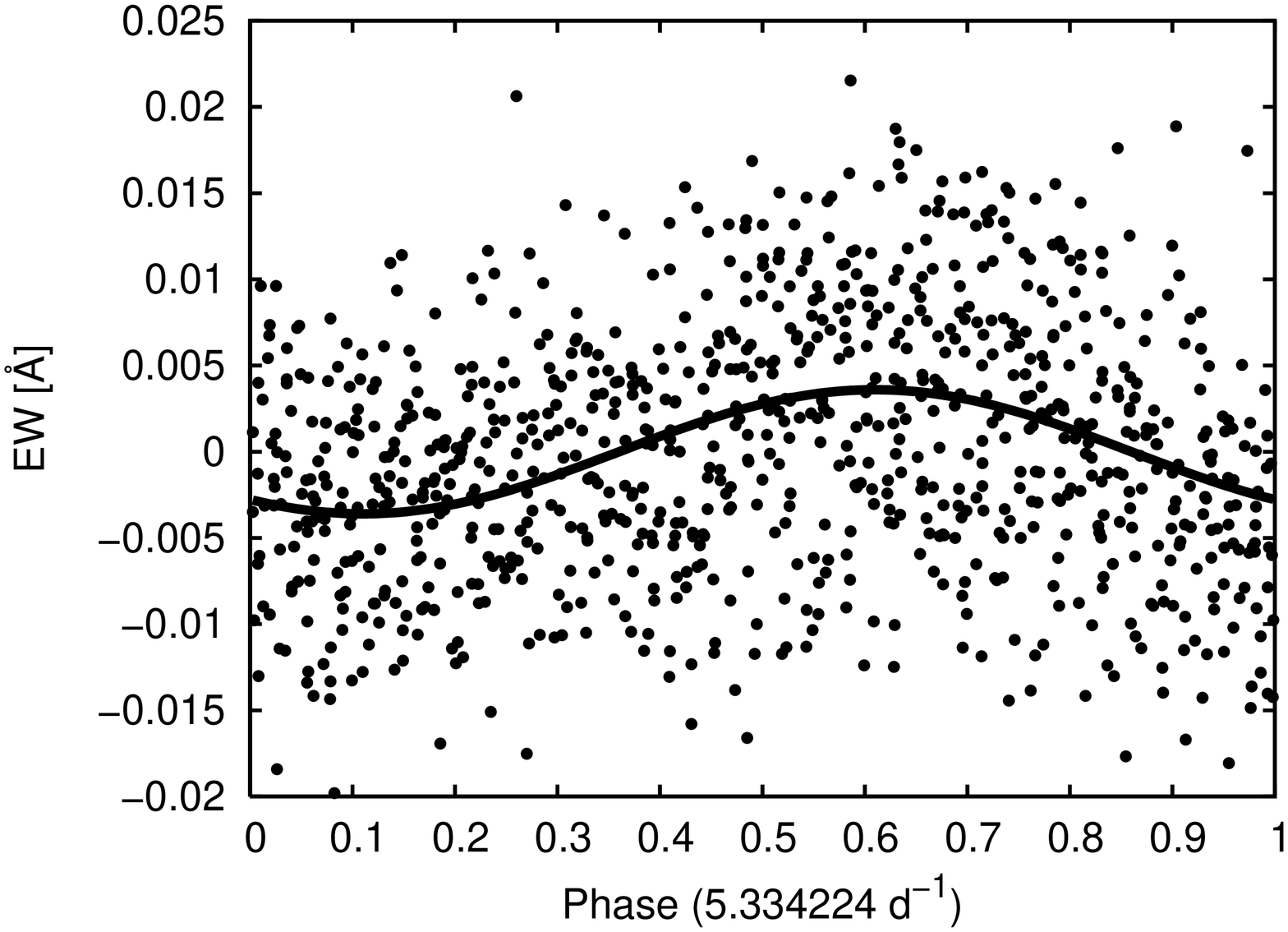}}}
\caption{Phase diagrams of the equivalent width of the \SiIII $\lambda 4553$
\A line, for $f_1$ (top panel), and for $f_2$ (bottom panel) after
prewhitening with $f_1$ .}
\label{FIG:EW_phase}
\end{center}
\end{figure}

The equivalent width (EW) of the silicon line of 12 Lac clearly varies with
the two dominant modes $f_1$ and $f_2$. The sum frequency $f_1+f_2$ is even
present. Fig.\,\ref{FIG:EW_phase} represents phase diagrams for both
frequencies. Such a strong EW variability of $9 \%$ is generally not observed
in pulsating B-type stars. Indeed, typical relative EW peak-to-peak amplitudes
are of the order of a few percent
\citep{2002A&A...385..572D,2002A&A...393..965D}.  However, for several $\beta$
Cephei stars, one clear sinusoidal variation is observed in their EW, with the
same frequency as in the radial velocity. The known cases are targets
pulsating multiperiodically with a high-amplitude dominant radial mode, such
as $\delta$ Ceti \citep{1992A&A...266..294A}, $\nu$ Eridani
\citep{2004MNRAS.347..463A} and V1449~Aql (Briquet et al., in preparation). In
the case of 12 Lac, we note that the EW signal is multiperiodic with a
dominant $(1,1)$-mode, followed by a radial mode of lower amplitude (see next
section).  For the four strongest modes, the frequency values derived from our
spectroscopic measurements are exactly the same, within the errors, as the
photometric ones by \citet{2006MNRAS.365..327H}. For the modes with radial
velocity amplitudes typically lower than 1 \kms, there are differences between
values derived from both kinds of data, which might have some influence on the
mode identification outcome. 
 Because the number of observations as well as the
time span of the datasets in the literature are much larger than the dedicated
spectroscopic ones, the literature frequency values are expected to be more
accurate and hence we adopt them. 

\section{Mode identification} \label{SEC:MODEID}

\subsection{The methods used}

We used two different and independent methods to identify the modes, namely
the moment method \citep{2003A&A...398..687B} and of the Fourier
parameter fit method \citep[FPF method,][]{2006A&A...455..227Z}.

The basic principles of both methods are the following. With the moment method
the wavenumbers $(\l,\m)$ and the other continuous velocity parameters are
determined in such a way that the theoretically computed first three moment
variations of a line profile best fit the observed ones. 
In the FPF method, for every detected pulsation frequency, the zero point,
amplitude, and phase are computed for every wavelength bin across the profile
by a multi-periodic least-square fit. Subsequently the zero-point profile, the
amplitude and the phase values across the profile are fitted with theoretical
values derived from synthetic line profiles.
The first technique has the advantage to be less
computationally demanding, allowing to test a huge grid for $(\ell,m)$ and for
the other parameters.  Moreover, the modes are determined simultaneously by
fitting a multiperiodic signal taking into account the coupling terms
appearing in the second and third moments. For the second technique, the
fitting is carried out by applying genetic optimization routines in a large
parameter space and mono-mode fits are used in order to speed up the
computations. Once the best solutions are constrained, a multi-mode fit is
performed. The strength of the FPF method is its ability to estimate the
significance of the derived mode parameters by means of a $\chi^2$ test 
\citep[all $\chi^2$-values listed in the text are reduced $\chi^2$ values, 
see][]{famias}. 

The FPF method is not optimally suited to study high-amplitude modes, but
as described further, the application of the FPF method for 12 Lac proved to be
succesful.
As shown by \citet{2004CoAst.144....5Z} and by \citet{2006A&A...455..227Z},
both mode identification techniques constrain the value of the
azimuthal number $\m$ better than the degree $\ell$ for low \vsini.  
Since the photometric data
provided us with unambiguous \l-values for the 5 main modes
\citep{2006MNRAS.365..327H}, a valid strategy is to make use of
spectroscopy to identify the values of \m, adopting the \l-values obtained
from photometry. Such a methodology already proved to be successful for two
other $\beta$ Cephei stars \citep{2005MNRAS.362..619B,2006A&A...459..589M}.
Mode identification results have been obtained with the software package
FAMIAS\footnote{FAMIAS has been developed in the framework of the FP6 European
Coordination Action HELAS -
\url{http://www.helas-eu.org/}.}
\citep{famias}.

In what follows we present in detail our mode identification by the FPF
method.  The moment method gave compatible results, but did not give
additional constraints on the $(\ell,m)$ than the ones we present here. Both
techniques found the same \m-values for the three main modes, giving us much
confidence in our outcome. For the fourth mode, the moment method derived the
sign of \m in agreement with the FPF method, which could further constrain its
value.

\subsection{Derivation of the wavenumbers}

For the FPF method, a sophisticated theoretical formalism for the modelling of
the displacement field is used and includes,
for instance, first-order effects of the Coriolis force as well as temperature
variations of the stellar atmosphere. 
 We refer to
\citet{2006A&A...455..227Z,famias}
for details on the computation of the synthetic line-profile variations. The
fixed parameters are the mass and the radius. 
When taking into account temperature effects, the
effective temperature $T_{\rm{eff}}$ and the logarithm of the gravity $\log g$
have also to be fixed. For 12 Lac, we adopted the
values $M
=13.5\, M_\odot$ \citep{2006MNRAS.365..327H}, $R =6.9\,R_\odot$
\citep{2006MNRAS.365..327H}, $T_{\rm{eff}} = 24500\pm1000\,\K$
\citep{2006A&A...457..651M}, $\log g = 3.65 \pm 0.15$
\citep{2006A&A...457..651M}. 
We note that the mode identification results are robust when using different
values for these parameters, within the errors.

\begin{figure}
\begin{center}
\rotatebox{-90}{\resizebox{!}{\columnwidth}{\includegraphics{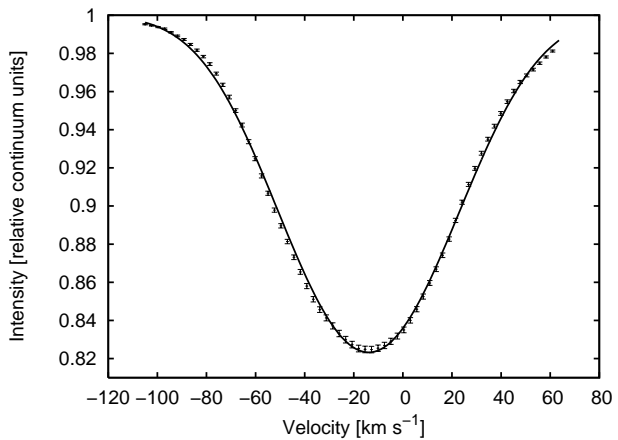}}}
\caption{Observed zero-point profile (points)  and best fit when taking into
account the four main modes (solid line) by
means of a synthetic rotationally broadened profile ($v \sin i=36.7\,\kms$,
W=$15.3\,\kms$, $\sigma=22.1\,\kms$).  
} 
\label{FIG:ZEROFPF} 
\end{center}
\end{figure}

In a first step we identified the 3 main modes ($f_{1,3,4}$) by fitting
mono-mode Fourier parameters, which are computed from synthetic line profiles
evenly sampled over one pulsation cycle \citep[for more explanations
see][]{2006A&A...455..227Z}. Moreover, we neglected the equivalent width
variations of the spectral line due to temperature variations. In order to
restrict the parameter space, we estimate roughly the values of the projected
rotational velocity $v \sin i$, the width of the intrinsic gaussian profile
$\sigma$, and the equivalent width $W$ from a least-squares fit of a
rotationally broadened synthetic profile to the zero-point profile. 
This gives only a crude estimate of these three
parameters, because the pulsational broadening is ignored in this fit, but it
serves the purpose of defining our appropriate grid. Moreover \vsini will
be overestimated. When taken into account the pulsational broadening caused by
the four main modes we acquire a \vsini of 36.7\,\kms (see 
Fig.\,\ref{FIG:ZEROFPF}).
A genetic optimization with
the following free parameters was afterwards adopted: $\ell \in [0,4]$ with a
step of 1, $m \in [-\ell,\ell]$ with a step of 1, the surface
velocity amplitude $a \in [10,110]\,\kms$ with a step of 1\,\kms, the stellar
inclination angle $i \in [5,85]\,\deg$ with a step of $5\,\deg$, $v \sin i \in
[20,48]\,\kms$ with a step of 0.5\,\kms and $\sigma \in [10,30]\,\kms$ 
with a step of 0.1\,\kms. 
We fixed \ft as a
radial mode and only fitted the continuous parameters for that mode. 

\begin{table*}
\centering
\begin{minipage}{180mm}
  \caption{Mode parameters derived from the FPF mono-mode method. 
For each pulsation mode, the five best solutions are shown together with the
second best $(\ell,m)$ identifications. For $f_5$ we list the different
solutions for \m. $a_s$ is the
surface velocity amplitude (\kms); $i$ is the stellar 
inclination angle in
degrees; $v\sin i$ is the projected rotational velocity, $\sigma$ is the width
of the intrinsic profile, both expressed in \kms. Due to spherical symmetry we
have not indicated the inclination angle for radial modes. The $95\%$
significance limit is at $\chi^2=1.15$.
}
\label{TAB:FPF}
\begin{center}
  \begin{tabular}{@{}ccccccc@{\hspace{1.5cm}}ccccccc@{}}
\hline \hline
\multicolumn{7}{c}{\fo = 5.178964\,\cd} & \multicolumn{7}{c}{\fth =
5.066316\,\cd}\\
$\chi^2$&\l&\m&$a_s$&$i$&$v\sin i$&$\sigma$ 
&$\chi^2$&\l&\m&$a_s$&$i$&$v\sin i$&$\sigma$ \\
&&&[\kms]&[\deg]&[\kms]&[\kms]&&&&[\kms]&[\deg]&[\kms]&[\kms]\\
\hline
15.29&1&1&68.7&74.6&46.47&21.00           &10.34&1&0&24.8&59.1&42.69&26.23\\
15.30&1&1&64.9&85.0&38.15&27.82           &10.38&1&0&24.2&59.2&42.70&26.23\\
15.31&1&1&48.2&72.1&43.72&21.64           &10.39&1&0&25.0&51.5&42.38&26.24\\
15.32&1&1&72.3&51.4&36.03&28.62           &10.40&1&0&26.4&48.9&42.38&26.35\\
15.33&1&1&50.9&59.2&42.00&22.58        &10.41&1&0&25.7&59.2&43.02&26.41\\[1ex]
31.62&1&0&68.4&41.2&30.79&26.66           &25.99&1&1&18.0&74.6&43.65&26.70\\
\hline
\multicolumn{7}{c}{\ft =  5.3324224\,\cd} & \multicolumn{7}{c}{\ff =
5.490133\,\cd}\\
$\chi^2$&\l&\m&$a_s$&$i$&$v\sin i$&$\sigma$ 
&$\chi^2$&\l&\m&$a_s$&$i$&$v\sin i$&$\sigma$ \\
&&&[\kms]&[\deg]&[\kms]&[\kms]&&&&[\kms]&[\deg]&[\kms]&[\kms]\\
\hline

12.36&0&0&41.6& -&43.65&25.17           &7.73&2&1&26.2&15.3&40.15&27.23\\
12.83&0&0&37.0& -&36.66&28.11           &7.76&2&1&26.2&12.7&40.16&27.17\\
13.07&0&0&42.6& -&43.33&25.00           &7.77&2&1&26.2&12.7&40.79&26.76\\
13.53&0&0&33.4& -&40.47&26.41           &7.79&2&1&27.1&12.7&40.79&27.11\\
13.68&0&0&33.3& -&44.60&24.70        &7.83&2&1&24.2&12.8&42.38&26.35\\[1ex]
     & & &    &  &     &               &43.68&2&2&23.7&82.4&41.76&26.26\\
%
\hline
\end{tabular}

\end{center}
\end{minipage}
\end{table*}

In Table \ref{TAB:FPF}, the 5 best solutions from a fit to the Fourier
parameters are listed for each frequency. The 95\%-confidence limit of
$\chi^2$ is 1.15. The most probable solutions are $(\l_1,\m_1)=(1,1)$,
$(\l_3,\m_3)=(1,0)$ and $(\l_4,\m_4)=(2,1)$ (where a positive \m-value denotes
a prograde mode).  We point out that the agreement between theory and
observations is not perfect. More precisely, the observed asymmetry with
respect to the centroid of the amplitudes across the line of the dominant
modes is not completely reproduced by our model.  
Such an asymmetry can be caused by EW variations of the intrinsic line
profile due to local temperature variations \citep{1999A&A...342..453S}. We
considered EW variations in our fitting procedure but we did not obtain an
improved solution compared to the case of a constant EW.

\begin{figure*}
\centering
\begin{center}
 \begin{tabular}{cc}
\vspace*{-2mm}
\resizebox{0.45\textwidth}{!}{\includegraphics{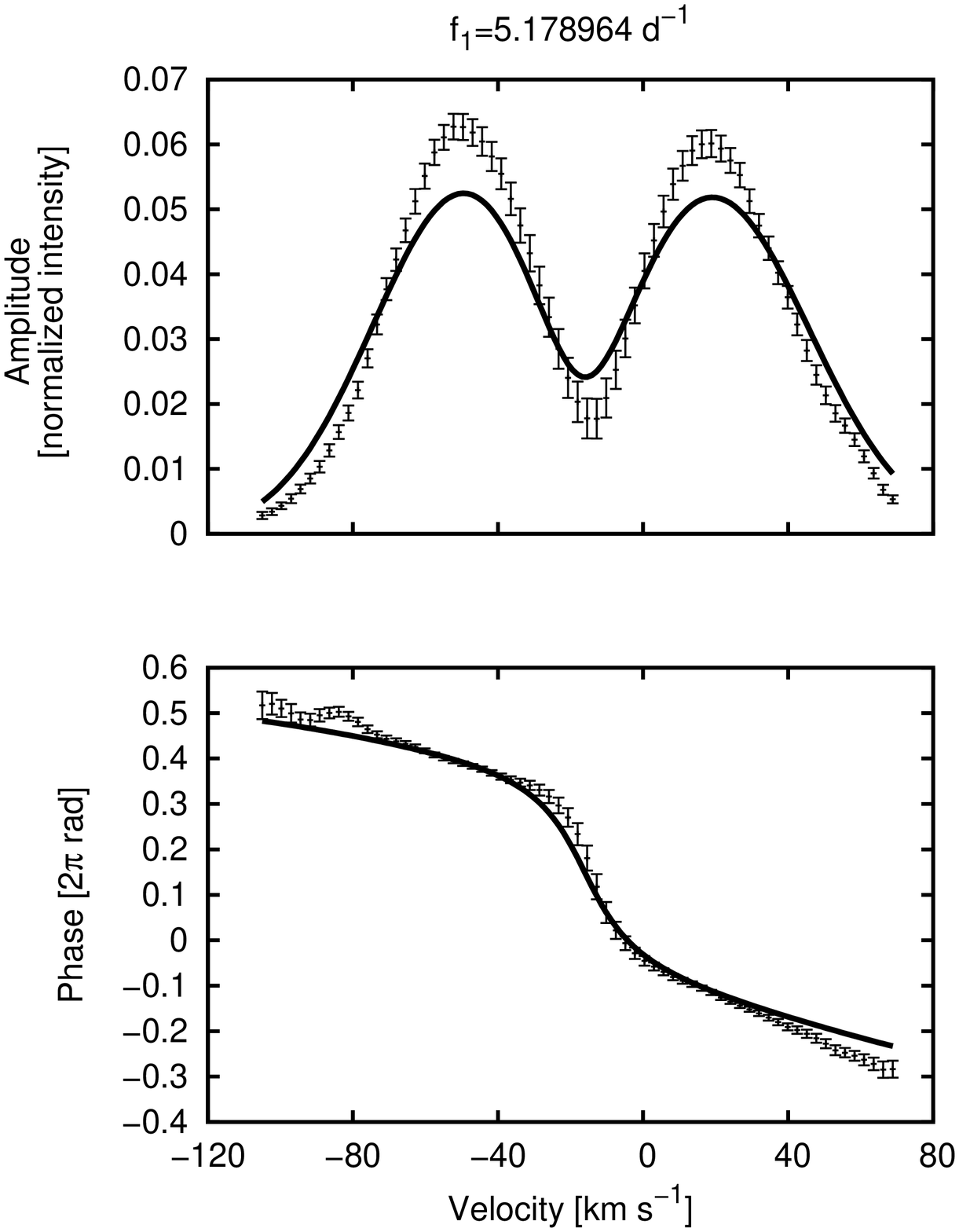}}&
\resizebox{0.45\textwidth}{!}{\includegraphics{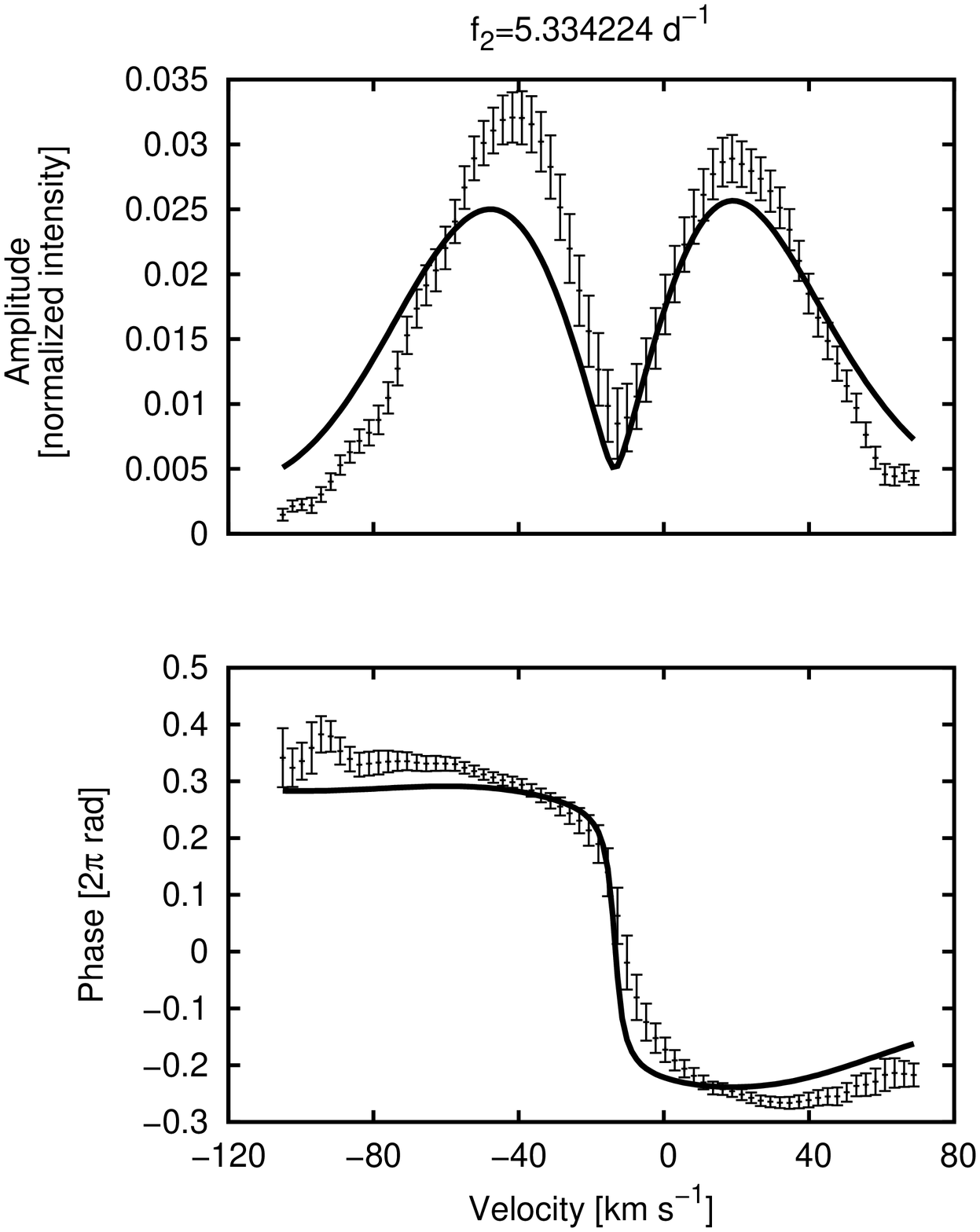}}\\
\vspace*{-2mm}
\resizebox{0.45\textwidth}{!}{\includegraphics{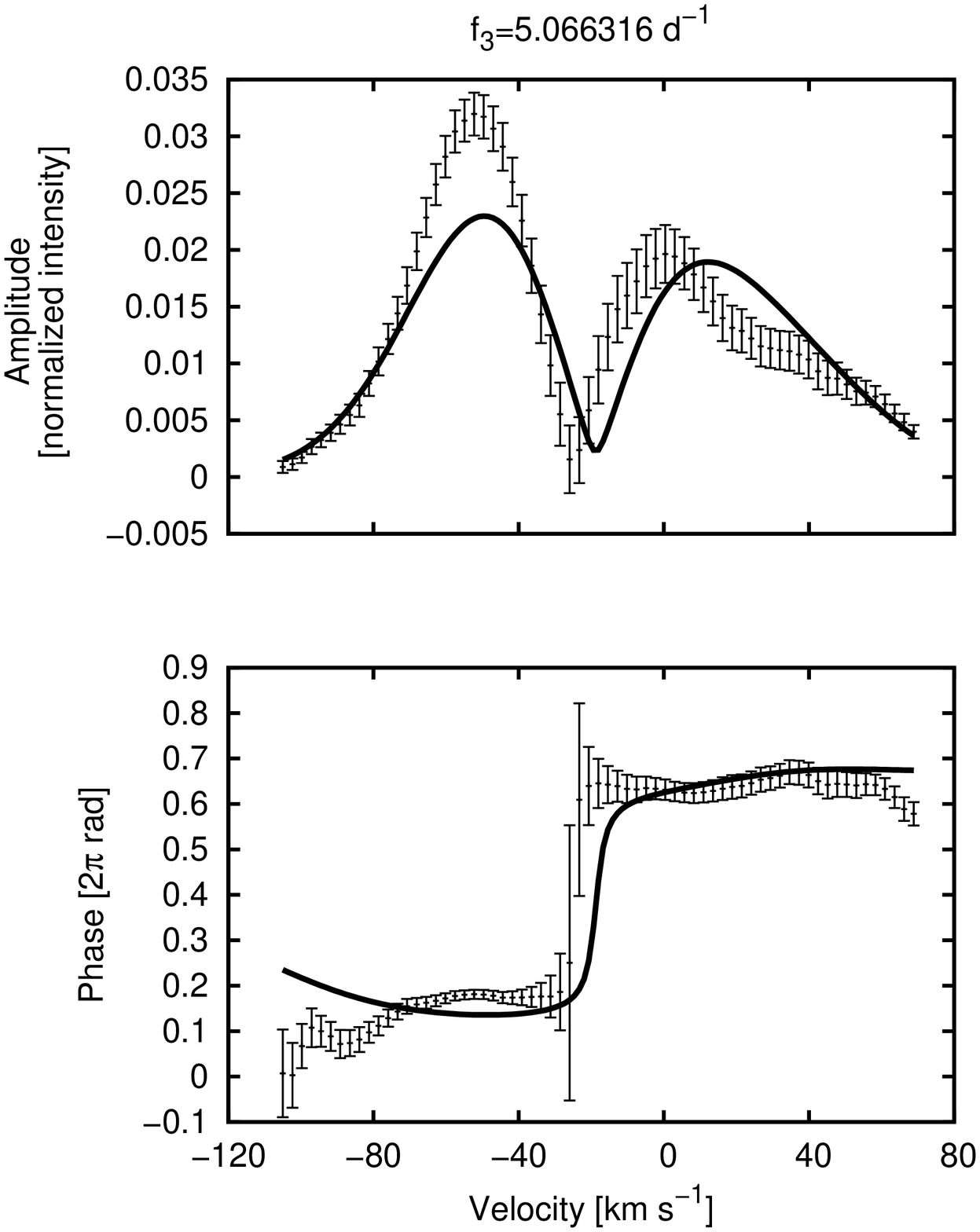}}&
\resizebox{0.45\textwidth}{!}{\includegraphics{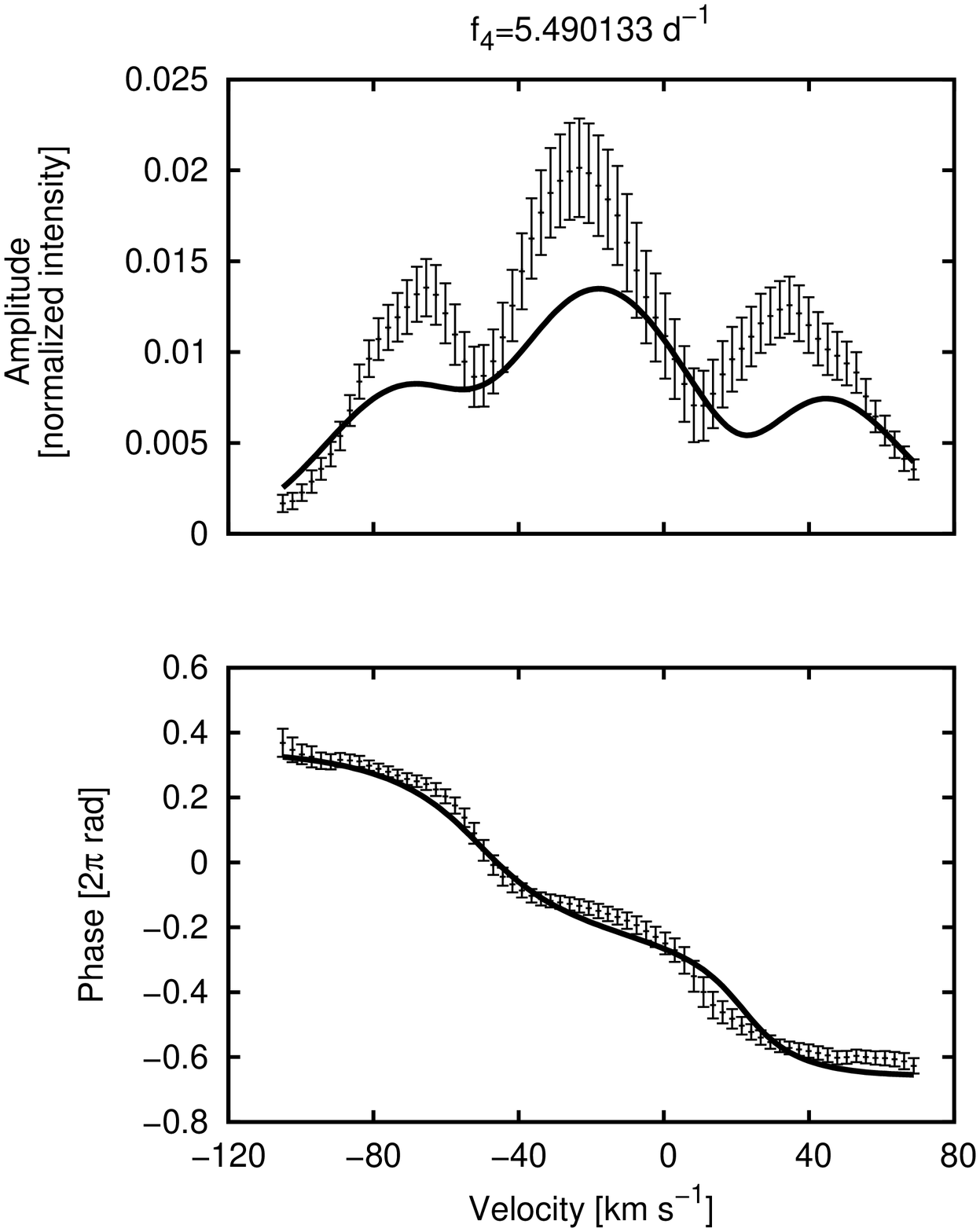}}\\
\end{tabular}
\end{center}
\vspace*{-5mm}
\caption{Observed Fourier parameters (black points with error bars) across the
line profile together with the best fit for the four identified modes (full
line). The amplitudes are expressed in units of continuum and the phases in
$2\pi$ radians. }
\label{FIG:BESTMODELSFPF}
\end{figure*}

In a second step we attempted to improve the solutions by simultaneously fitting
 the four main modes and using all the original data points.  This however was
 computationally not possible. To restrict the computation time we took a subset
 of our complete time series. Our subset resulted in 388 spectra with a time
 span of 4 months.  We also fixed the wave numbers according to our mono-mode
 fit described above.  We succeeded in improving the surface velocity amplitudes
 together with the inclination, \vsini and the width of the intrinsic profile.
 The $\chi^2$ values are lower than in the mono-mode fit and we are able to
 reproduce part of the asymmetry in the amplitude across the line as explained
 above (see Table \ref{TAB:FPFSIM}).  Fig.\,\ref{FIG:BESTMODELSFPF} illustrates
 the best models for the amplitude and phase across the line profile for the
 four main modes.

\begin{figure*}
\centering
\begin{center}
 \begin{tabular}{ccc}
\vspace*{-2mm}
\resizebox{0.30\textwidth}{!}{\includegraphics{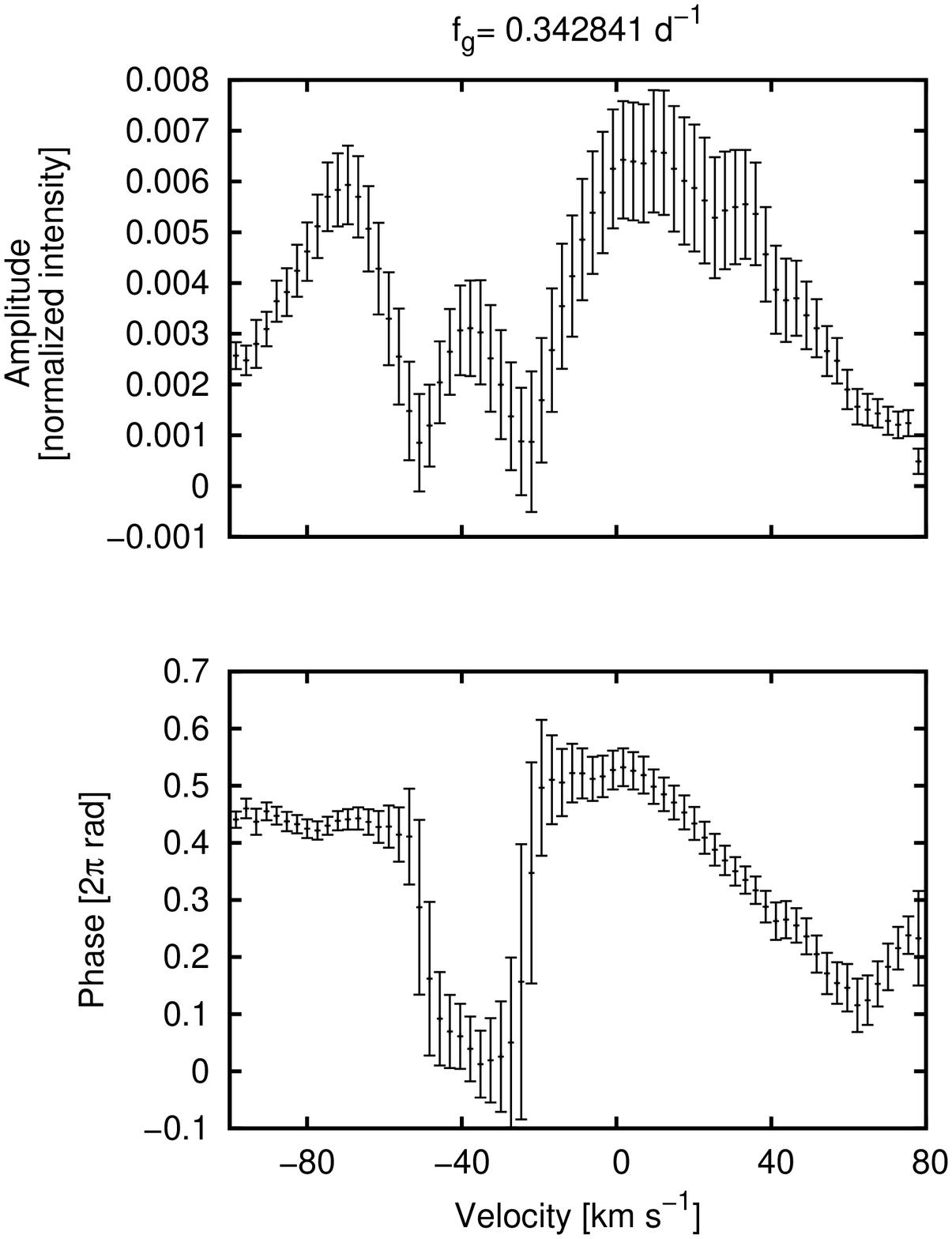}}&
\resizebox{0.30\textwidth}{!}{\includegraphics{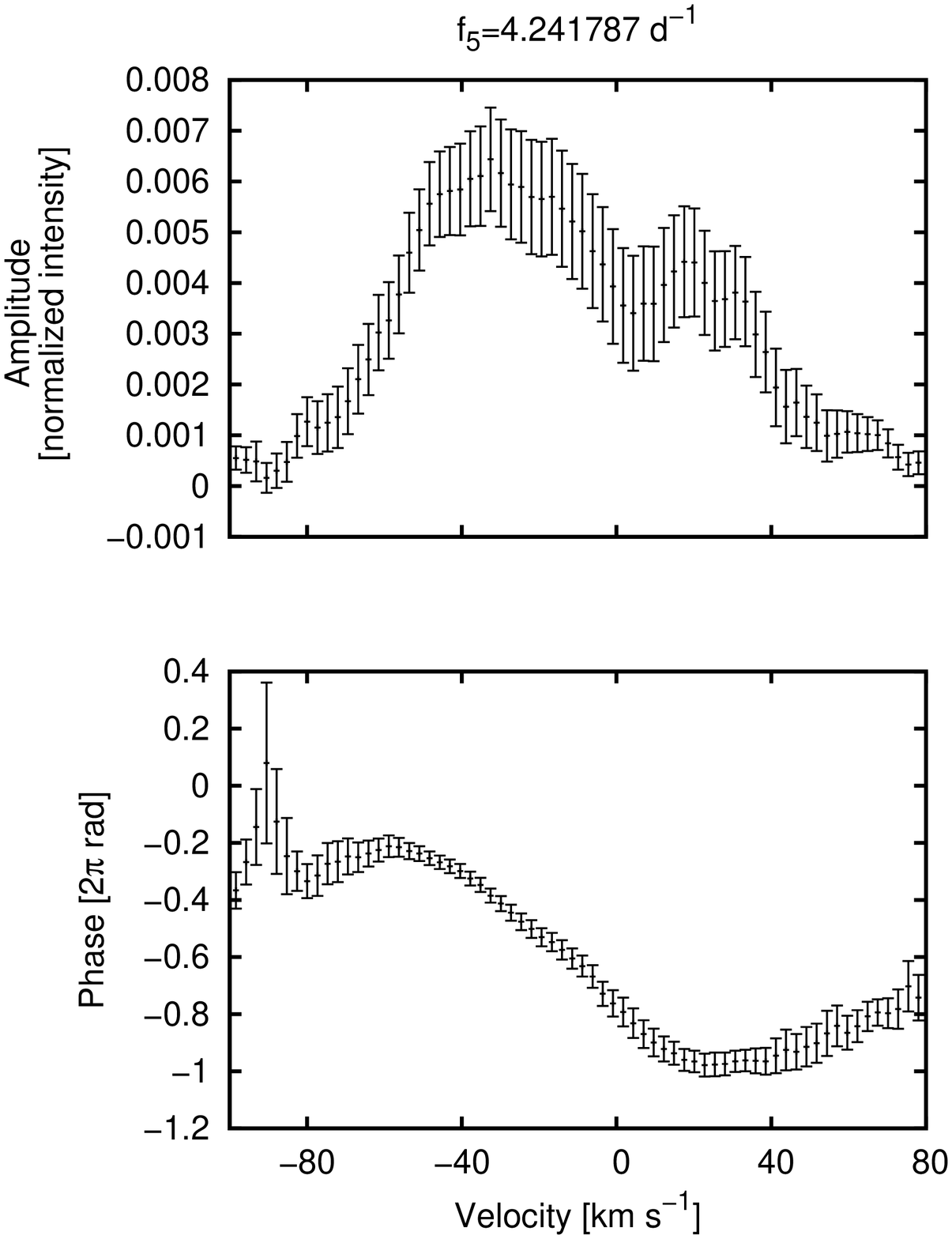}}&
\resizebox{0.30\textwidth}{!}{\includegraphics{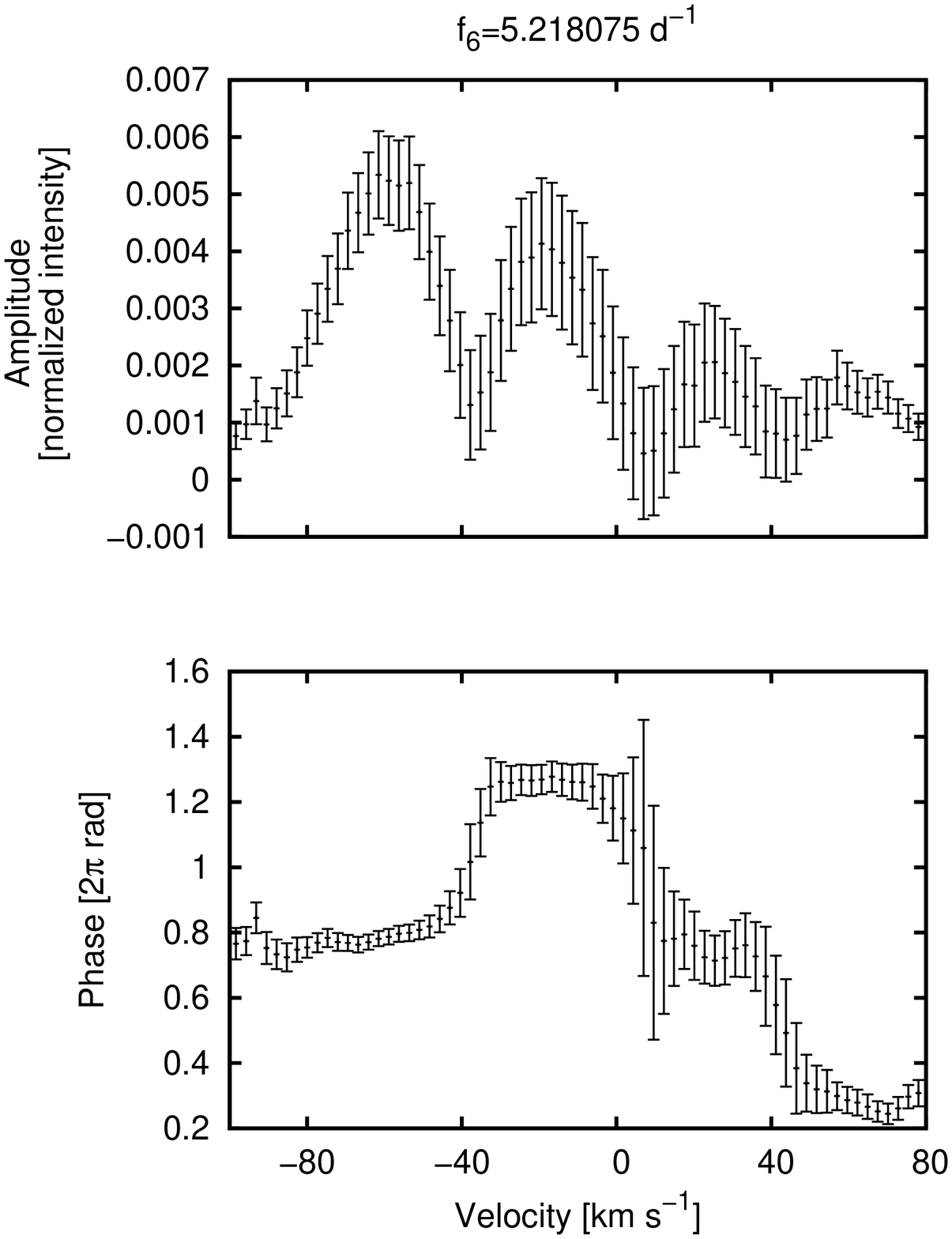}}\\
\vspace*{-2mm}
\resizebox{0.30\textwidth}{!}{\includegraphics{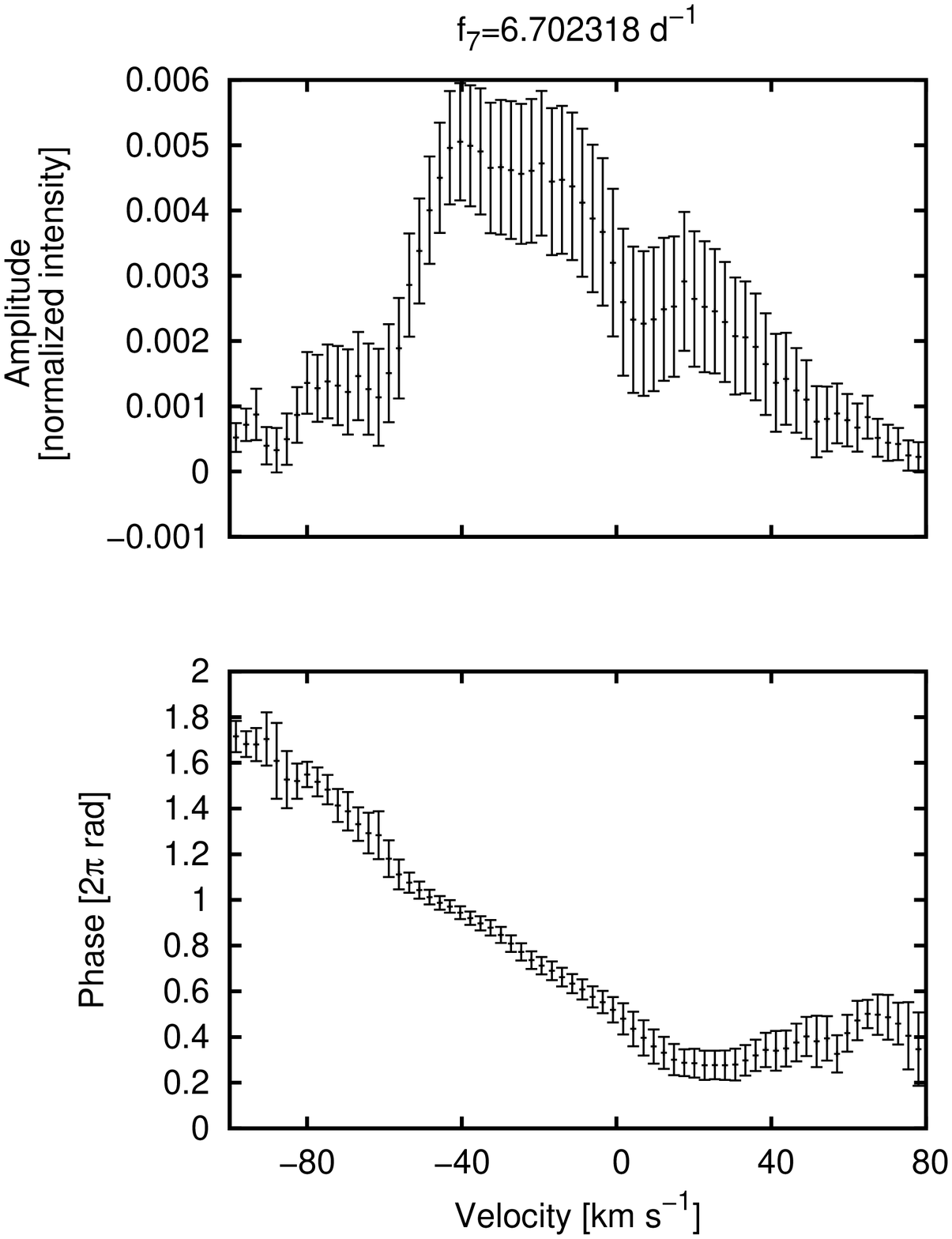}}&
\resizebox{0.30\textwidth}{!}{\includegraphics{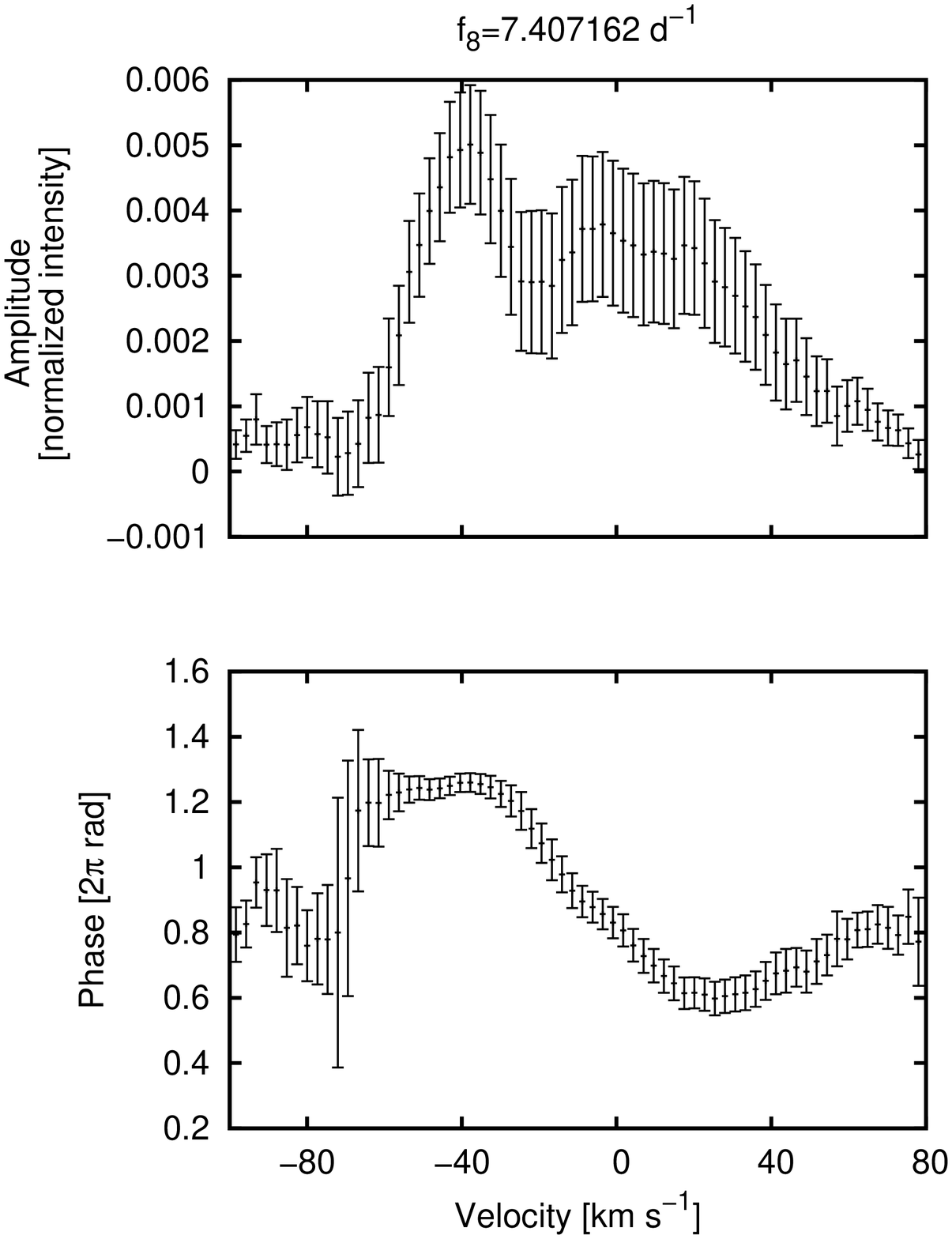}}&
\resizebox{0.30\textwidth}{!}{\includegraphics{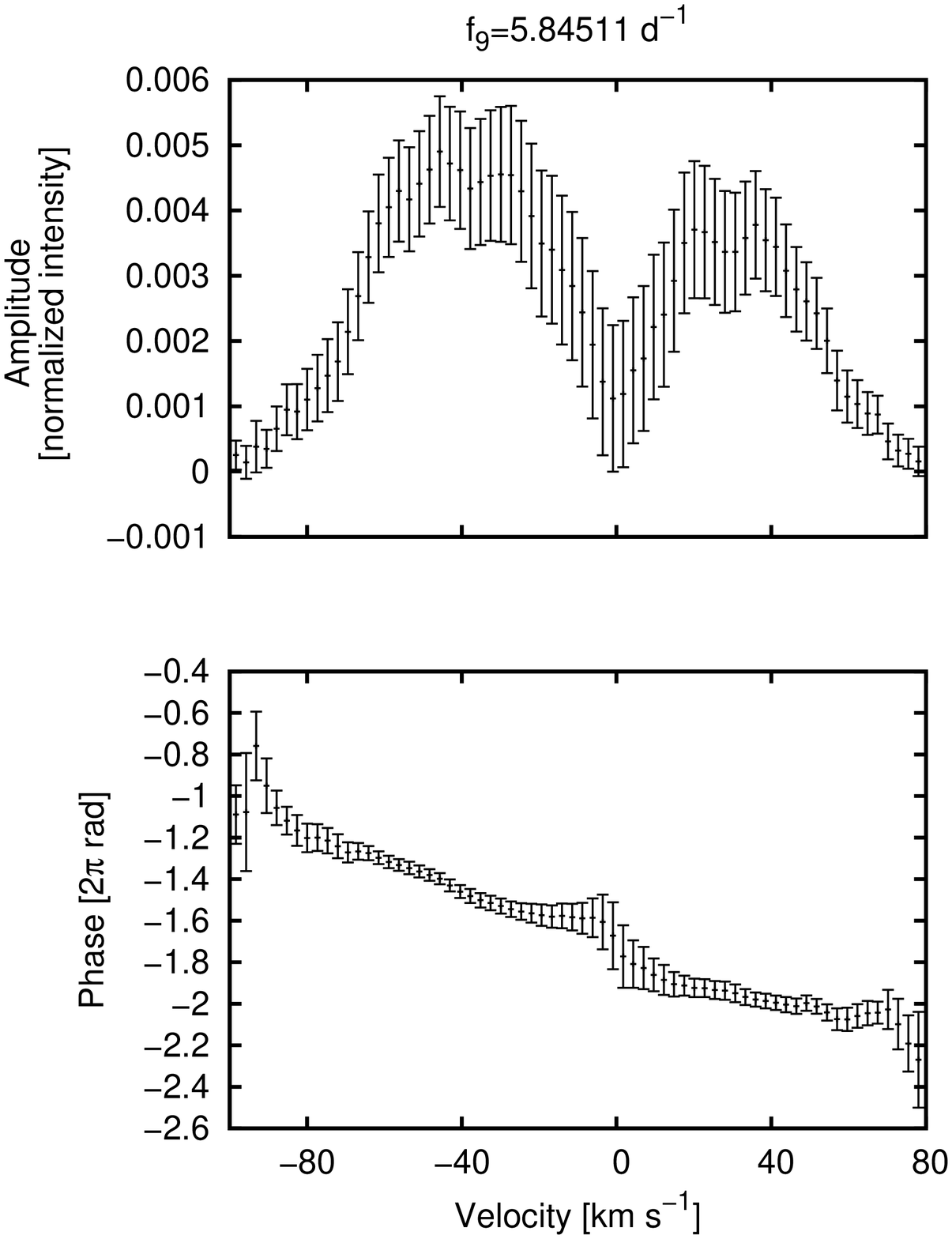}}\\
\end{tabular}
\end{center}
\vspace*{-5mm}
\caption{Observed Fourier parameters (black points with error bars) across the
line for the detected pulsation frequencies ($f_{g,5-9}$). The amplitudes are
expressed in units of continuum and the phase in $2\pi$ radians.
}
\label{FIG:RESTAMPPHASE}
\end{figure*}

Finally, we tried to identify the other low-amplitude modes with the same
methodology and by imposing the wavenumbers $(\ell,m)$ deduced for the 4 main
modes. Unfortunately, we could not obtain additional conclusions. Our failure is
not very surprising since these modes have amplitudes lower or of the same order
as the harmonics and combination frequencies also present in the signal while
non-linear terms are not taken into account in our line-profile modelling.
Fig.\,\ref{FIG:RESTAMPPHASE} depicts the amplitude and phase across the line for
these low amplitude modes.  Despite the fact that the phase across the profile
is well determined for these modes and seems to point to axisymmetric or
prograde modes, the $\chi^2$ values between the different options were not
discriminative so we do not use that information in the following.  In Table
\ref{TAB:LM}, we summarize the final outcome of both the photometric
\citep{2006MNRAS.365..327H} and spectroscopic mode identifications.

\begin{table}
\centering
\caption{Mode parameters derived from the FPF method
through a simultaneous fit of the four main modes. 
We used the same symbol conventions as in Table\,\ref{TAB:FPF}. The $95\%$
significance limit is at $\chi^2=1.31$.
}
\label{TAB:FPFSIM}
\begin{tabular}{@{}cccccccc@{}}
\hline\hline
\multicolumn{8}{c}{$f_{1;(1,1)}$, $f_{2;(0,0)}$, $f_{3;(1,0)}$, $f_{4;(2,1)}$}\\
$\chi^2$&$a_{s1}$&$a_{s2}$&$a_{s3}$&$a_{s4}$&$i$&$v\sin i$&$\sigma$\\
&\multicolumn{4}{c}{[\kms]}&[\deg]&[\kms]&[\kms]\\
\hline
7.26 & 91.6 &  54.9 &24.5 & 29.6 & 43.7 & 36.7 & 22.1 \\
7.34 & 91.6 &  59.3 &26.9 & 28.8 & 46.2 & 37.0 & 21.6 \\
7.51 & 98.4 &  57.4 &18.0 & 28.8 & 43.7 & 35.1 & 22.0 \\
7.55 & 83.8 &  51.1 &24.5 & 28.8 & 54.0 & 35.4 & 21.7 \\
7.64 & 90.6 &  50.4 &26.9 & 28.8 & 43.7 & 37.1 & 22.1 \\
\hline
\end{tabular}
\end{table}


\begin{table}
\centering
  \caption{Final results for the mode identifications of 12 Lac 
from our spectroscopic analysis together with the results from the 
photometric amplitude ratios
\citep{2006MNRAS.365..327H}.
}
\label{TAB:LM}
  \begin{tabular}{@{}lcccc}
\hline \hline
ID & Frequency &\multicolumn{2}{c}{\l} & \m  \\
& [\cd] & Spectr. & Phot. &     \\
\hline
\fo  &          5.178964   & \bf1     &  1  &  \bf 1    \\
\ft  &          5.334224   & \bf0     &  0   & \bf 0     \\
\fth &          5.066316   & \bf1     &  1   & \bf 0    \\
\ff  &          5.490133   & \bf2     &  2   & \bf 1    \\
$f_g$    &      0.342841   & -- &  1,2,4   &  --   \\
$f_5    $ &     4.241787   & --     &  2   &  $0,1,2$  \\
$f_6    $ &     5.218075   & --   &  2,4   &    --  \\
$f_7    $ &     6.702318   & --     &  1   &    --   \\
$f_8    $ &     7.407162   & --   &  1,2   &    --   \\
$f_9    $ &     5.84511   &  --   &  1,2   &    -- \\
$f_p    $ &     5.30912   &  --   &  1,2   &  -- \\
\hline
\end{tabular}
\end{table}


\subsection{Derivation of the surface equatorial rotational velocity and the inclination}

\begin{figure}
\rotatebox{0}{\resizebox{\columnwidth}{!}{\includegraphics{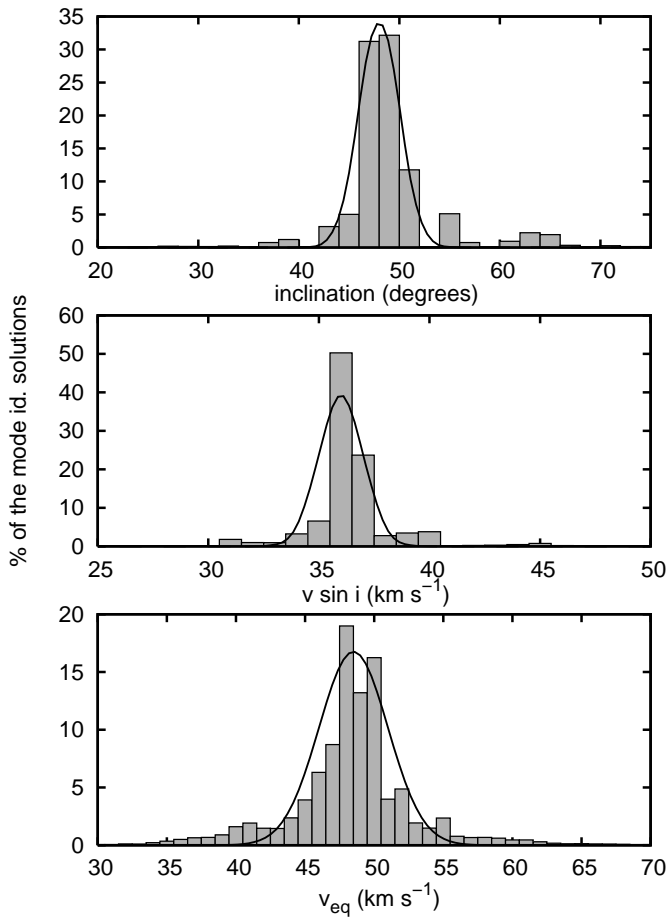}}}
\caption{Histograms for the inclination, projected rotation velocity and 
equatorial rotational velocity of the star derived from 
the FPF method.} 
\label{FIG:histogram_veq}
\end{figure}

From mode identification we can also estimate the inclination, \vsini and the
equatorial rotational velocity \veq. By using $\chi^2$ as a weight we can
construct histograms for $i$, \vsini and \veq. In
Fig.\,\ref{FIG:histogram_veq} we display these histograms. They are computed
by considering all the solutions of the FPF method with the correct wavenumbers
of the four dominant modes through a simultaneous fit,
and by giving each parameter ($i_k$, $v \sin i_k$ and
$v_{eq,k}$) its appropriate weight $w_k=\chi^2_0 / \chi^2_k$, where $\chi^2_0$
is the $\chi^2$-value for the best solution. By calculating a weighted mean
and standard deviation of the data, we get $i=48\pm2\,^{\circ}$, $\vsini=36\pm
2\,\kms$ and $\veq= 49 \pm 3\,\kms$. This leads to a surface rotational
frequency between 0.11 and 0.12\,\cd for the appropriate stellar models in
Table\,\ref{model_parameters}, which will be discussed further on. 
In the same way we constructed these histograms with the solutions from the
multi-mode fit of the moment method.  However, the moment method did not have
the ability to limit the range for $i$ and, by implication, leads to a less
trustworthy estimate of \veq. The reason for this is
probably that the moment method is based on integrated quantities over the
line profile, while in the FPF method we use all the information across the
line profile as a whole, which is more sensitive to the position of the nodal
lines across the stellar surface and thus to the stellar inclination.

\section{Modelling}

In this section, we check if state-of-the-art stellar models with standard
physics can account for the observed frequency spectrum together with the
derived wavenumbers $(\ell,m)$ for our studied star. By doing so, we can
constrain 12 Lac's model parameters, which are the mass, the central hydrogen
abundance and the core convective overshooting parameter. Such asteroseismic
constraints have recently been derived for a few $\beta$~Cephei stars: 16 Lac
\citep{2003A&A...406..287T}, V836~Cen
\citep{2003Sci...300.1926A,2004A&A...415..251D}, $\nu$~Eri
\citep{2004MNRAS.350.1022P,2004MNRAS.355..352A,2008MNRAS.385.2061D},
$\delta$~Ceti \citep{2006ApJ...642..470A}, $\beta$~CMa
\citep{2006A&A...459..589M}, $\theta$~Oph \citep{2007MNRAS.381.1482B}.
Finally, \citet{Ausseloosphd} and, more recently, \citet{2008MNRAS.385.2061D}
already made a comparison with stellar models for 12 Lac.  

\subsection{Numerical tools and physical inputs} In our work, we used the
following numerical tools and physical inputs. The stellar models for
 non-rotating 
stars were computed with the evolutionary code CL\'ES \citep[Code
Li\'egeois d'\'Evolution Stellaire,][]{2008Ap&SS.316...83S}.  We used the
OPAL2001 equation of state \citep{2002ApJ...576.1064R,1988ADNDT..40..283C},
with nuclear reaction rates from \citet{2004PLB..591..61C} for the
$^{14}$N$(p,\gamma)^{15}$\,O cross-section. Convective transport is treated by
using the classical Mixing Length Theory of convection
\citep{1958ZA.....46..108B}.  For the chemical composition, we used the solar
mixture from \citet{2005ASPC..336...25A}. We used OP opacity tables
\citep{2005MNRAS.362L...1S} computed for this latter mixture. These tables are
completed at $\log T < 4.1$ with the low temperature tables of
\citet{2005ApJ...623..585F}. 

Our choice for the metal mixture and the opacities is justified by the
following arguments. First, \citet{2006A&A...457..651M} found that the
abundances of He, C, N, O, Mg, Al, Si, S and Fe are, within the errors, in
good agreement with the solar values derived from time-dependent,
three-dimensional hydrodynamical models of \citet[][and references
therein]{2005ASPC..336...25A}.  Secondly, \citet{Ausseloosphd} performed a
preliminary seismic study of 12 Lac and was confronted with the problem that,
when using OPAL opacities, not all observed pulsation modes can be excited for
realistic values of the metallicity and core overshooting parameter. This
excitation problem is encountered with the \citet{1993A&A...271..587G} but
also with the \citet{2005ASPC..336...25A} mixture,  even for large values of
$Z$ (typically larger than 0.016). If we consider values of $Z$ smaller than
0.016, which are more representative of the star, the problem is obviously
increased. As discussed in \citet{2007MNRAS.381.1482B} for $\theta$~Oph, modes
that are stable with OPAL opacities
\citep{1992ApJS...79..507R,2002ApJ...576.1064R} can be predicted to be
unstable with OP opacities. The reason is that in a typical $\beta$~Cephei
model the OP opacity is 25\% larger than  the one for OPAL in the region where
the driving of pulsation modes occurs. We consequently wish to check if the
use of OP opacity tables improves the situation in the case of the modelling
of 12 Lac.  We refer to \citet{2007MNRAS.375L..21M} for a detailed discussion
on the implication of the adopted opacity tables and metal mixtures on the
excitation of pulsation modes in the case of B-type stars.  

For each stellar model, we calculated the theoretical frequency spectrum of
low-order $p$- and $g$-modes with a degree of the oscillation up to $\ell = 4$
using a standard adiabatic code 
for non-rotating stellar models
\citep{2008Ap&SS.316..149S}, which is much
faster than a non-adiabatic code but leads to the same theoretical pulsation
frequencies within the adopted precision of the fit, which was 10$^{-3}$
d$^{-1}$. Once the models fitting the observed modes are selected, we checked
the excitation of the pulsation modes with the linear non-adiabatic code 
{\sc MAD}
developed by \citet{2001A&A...366..166D}.
 
\subsection{Seismic analysis} 

Besides the pulsation characteristics summarized in Table\,\ref{TAB:LM}, we
also use $T_{\rm{eff}}$, $\log g$ and the metallicity Z as additional
observational constraints. Geneva photometry of 12 Lac provides $T_{\rm{eff}}$
= 23500$\pm$700 K and $\log g$ = 3.4$\pm$0.4 \citep{2006MNRAS.365..327H}. From
IUE spectra, \citet{2005A&A...433..659N} derived $T_{\rm{eff}}$ =
23600$\pm$1100 K and $\log g = 3.65\pm0.15$.  \citet{2006A&A...457..651M}
obtained $T_{\rm{eff}}$ = 24 500$\pm$1000 K and $\log g$ = 3.65$\pm$0.1 from
optical spectra. Using another method on the same spectra as
\citet{2006A&A...457..651M}, \citet{Lefever} deduced $T_{\rm{eff}}$ =
24000$\pm$1000 K and $\log g$ = 3.60$\pm$0.1.  \citet{2006A&A...457..651M}
also determined a metallicity $Z$ = 0.0089$\pm$0.0018.

Stellar models are parametrized by the initial hydrogen abundance $X$, the
core convective overshooting parameter $\alpha_{\rm{ov}}$, the metallicity $Z$
and the mass $M$. We adopted $X=0.72$ but using a different value for $X$ in a
reasonable range does not change the conclusions.

We considered two values for $Z$, i.e. 0.010 and 0.015. The first chosen value
corresponds to a metallicity typical of B dwarfs in the solar neighbourhood,
also found for 12 Lac within the errors \citep[see
above,][]{2006A&A...457..651M}. The second choice corresponds to the solar
composition and the reason for also adopting this value is the following. The
star being a relatively slow rotator and its surface convective zone being
very thin, diffusion mechanisms can alter the photospheric abundances, leading
to $Z$ smaller at the surface compared to the interior of the star. However,
since these effects are limited to the very surperficial layers, including a
diffusion treatment or not does not affect the derived stellar parameters. It
has been carefully checked in the case of $\theta$~Oph. We refer to
\citet{2007MNRAS.381.1482B} for a deeper discussion. For simplicity, we
computed 
non-rotating 
stellar models without taking into account diffusion, but we also
considered a $Z$ higher than what is derived from spectra.

Several values for $\alpha_{\rm{ov}}$ in the range [0.0-0.5] were tested. The
reason is that recent asteroseismic modelling of several $\beta$~Cephei stars
(listed above) showed the occurrence of core overshooting with
$\alpha_{\rm{ov}}$ in the range [0.05-0.25] for all studied stars except
$\theta$~Oph for which an even higher value of $\sim 0.45$ was determined.
Moreover, other works corroborate the necessity to include core overshooting
in modelling of massive B-type stars. For instance,
\citet{2000ApJ...543..395D} derived a value of about 0.45 by means of 2D
hydrodynamic simulations of zero-age main-sequence convective cores. We also
mention that, with his study of 13 detached double-lined eclipsing binaries,
\citet{2007A&A...475.1019C} found that models with core overshooting of
$\alpha_{\rm ov}\sim 0.2$ are needed to analyse $\sim$10 $M_\odot$ stars.

The observed radial mode with $f_2 = 5.334224$\,\cd and the $\ell = 1$ zonal
mode with $f_3 = 5.066316$\,\cd were selected as the first two frequencies to
be confronted with the theoretically predicted frequencies. 
This was an obvious choice since these are 
zonal modes $(\m=0)$ and thus we do not need any information about the
equatorial frequency of rotation to fit them with the model frequencies. 
We considered two
different scenarios: $f_2$ being either the radial fundamental mode or the
first overtone.

The theoretical frequency spectrum with $f_2$ being the radial fundamental
mode is globally compatible with the observed frequencies of 12 Lac. However,
all such models fail to reproduce the $\ell = 1$ mode with $f_7 = 6.702318$
\cd. If $f_2$ corresponds to a radial fundamental mode, $f_7$ can only be an
$\ell = 2$ mode according to theoretical models, while the observed
photometric amplitude ratios for this mode  point out an $\ell = 1$
\citep{2006MNRAS.365..327H}. Unfortunately, we
cannot confirm this strong constraint since our spectroscopic mode
identification was not conclusive for this mode. However, stellar models with $f_2$
corresponding to the radial first overtone can account for a $\ell = 1$ mode
with $f_7$. We consequently favour a radial order $n=2$ for $f_2$. 
An additional strong argument to favour $f_2$ as the radial first overtone is
the following.  
The Ledoux rotational splitting is defined by
$f_{\rm Ledoux}=m\beta_{nl}\bar{f}_{\rm rot}$, where $\beta_{nl}$ is a structure constant
depending on the stellar model, and originates from the
frequency splitting defined as
\begin{equation}\label{eq:split}
f_m=f_0+m\beta_{nl}\bar{f}_{\rm rot}=\nu_0+m\int_0^R K(r)
\frac{\Omega(r)}{2\pi}\frac{{\rm d} r}{R},
\end{equation}
where $f_m$ is the cyclic frequency of a mode of azimuthal-order $m$, 
$\Omega(r)$ is the angular velocity, and $K(r)$ is the rotational kernel
\citep[for more information, see][]{2008Ap&SS.316..149S}.  
With $f_2$ as fundamental, the difference between the
frequency $f_5 = 4.241787$\,\cd and the nearest quadrupole mode (at
$\sim$3.8\,\cd) is too large to be explained by the Ledoux rotational
splitting and thus $f_5$ cannot be fitted.  
With $f_2$ as first
overtone, $f_5$ is easily identified as the $g_2$ mode with $(\ell,m)=(2,1)$ 
as
shown below (see also Fig.\,\ref{freq_spectrum}).  
With $f_2$ being the first overtone, $f_3$ is
identified as $p_1$.

\begin{figure}
\begin{center}
\rotatebox{0}{\resizebox{\columnwidth}{!}{\includegraphics{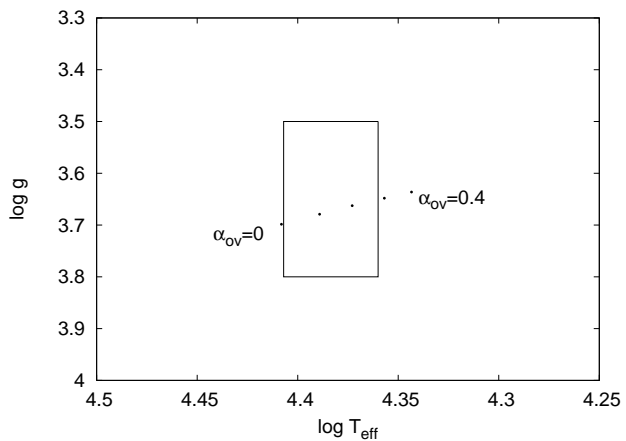}}}
\caption{The error box represents the position of 12 Lac in the
$\log (T_{\rm{eff}})-\log g$ diagram. The positions of the models which fit
exactly $f_2$ (being the first overtone) and $f_3$  are also shown for different $\alpha_{\rm{ov}}$ values. 
 } 
\label{hr}
\end{center}
\end{figure}

The couple $(X,Z)$ being fixed, the fitting of two frequencies suffices to
derive the stellar age and stellar mass for adopted $\alpha_{\rm{ov}}$. The
positions in the $\log (T_{\rm{eff}})-\log g$ diagram of the models matching
$f_2$ and $f_3$ for different values of $\alpha_{\rm{ov}}$ are shown in
Fig.\ref{hr}. The corresponding parameters are listed in Table\,\ref{model_parameters} for
$Z=0.015$. The mass decreases when the core overshooting parameter increases.
We also note the perfect agreement between the $\log g$ of the models and the
value deduced from spectroscopic observations. Moreover, the best agreement in
$T_{\rm{eff}}$ corresponds to models with $\alpha_{\rm{ov}}$ between 0.0 and
0.4 and a mass between 10.0 and 14.4 $M_\odot$. Note that a decrease in $Z$
implies an increase in mass and, considering a $Z$ of 0.010 instead of 0.015
increases the mass by 0.5 $M_\odot$ at most.

\begin{table}
\centering
\caption{Physical parameters of the models that match $f_2$ 
(being the first overtone) and $f_3$, with
$X=0.72$ and $Z=0.015$. $X_c$ is the central hydrogen abundance. The age
$\tau$ is expressed in million years.
  }

\begin{center}
\begin{tabular}{@{}c@{\hspace{2.0mm}}c@{\hspace{2.4mm}}c@{\hspace{2.4mm}}c@{\hspace{2.4mm}}c@{\hspace{2.4mm}}c@{\hspace{2.0mm}}c@{\hspace{2.0mm}}c@{}}
\hline\hline
$\alpha_{\rm{ov}}$ & M\,($M_\odot$) & $T_{\rm{eff}}$\,(K) & $\log g$ & $X_c$&
R\,$(R_\odot)$&$\log(L / L_\odot)$&$\tau\,(My)$\\
\hline
0.0 & 14.4 & 25600 & 3.70 & 0.13& 8.8 & 4.48&11   \\
0.1 & 13.1 & 24500 & 3.68 & 0.15& 8.6 & 4.38&13   \\
0.2 & 12.0 & 23600 & 3.66 & 0.17& 8.4 & 4.29&16   \\
0.3 & 11.0 & 22750 & 3.65 & 0.19& 8.2 & 4.21&20   \\
0.4 & 10.2 & 22000 & 3.64 & 0.21& 8.0 & 4.13&23   \\
\hline
\end{tabular}
\end{center}
\label{model_parameters}
\end{table}



\subsection{Radial order identifications}

\begin{table}
\centering
  \caption{The $\ell$, $n$ and \m identifications according to our 
mode identification and modelling with the radial mode as first overtone.
We use the usual convention
that negative $n$-values denote $g$-modes \citep[see
e.g.][]{1989nos..book.....U}. 
}
\label{order}
\begin{center}
  \begin{tabular}{ccccc}
\hline \hline
ID & Frequency &   $\ell$ & $n$  & \m \\
& [\cd] &    &    &\\
\hline
\fo  &          5.178964   &1 & 1          &1   \\
\ft  &          5.334224   &0 & 2          &0   \\
\fth &          5.066316   &1 & 1          &0   \\
\ff  &          5.490133   &2 & 0          &1   \\
$f_5    $ &     4.241787   &2 & -2         &1   \\
$f_6    $ &     5.218075   & 4 & -2 or -1 &   \\
$f_7    $ &     6.702318   & 1  &3         &   \\
$f_8    $ &     7.407162   & 2 &2         &   \\
$f_9    $ &     5.84511    & 1 &2         &  \\
$f_p    $ &     5.30912    & 2  &0        &0   \\
\hline
\end{tabular}
\end{center}
\end{table}


The frequency $f_1 = 5.178964$\,\cd can only belong to the $p_1$ triplet. The
frequency splitting $f_1 - f_3$ is 0.1126\,\cd. According to
the models, $\beta_{nl}=0.5$ for this mode.
This yields an averaged 
frequency of rotation $\bar{f}_{\rm rot}=(f_1-f_3)/m\beta_{nl}$. 
This averaged rotation frequency leads to a rotational velocity 
twice as high as the
surface value estimated from our spectroscopic observations (with
corresponding velocities of 100\,\kms 
versus 49\,\kms). 
We thus confirm that 12 Lac must rotate more rapidly in its inner parts 
than at the surface, as already emphasized by \citet{2008MNRAS.385.2061D}. 

\begin{figure*}
\begin{center}
\rotatebox{-90}{\resizebox{!}{\hsize}{\includegraphics{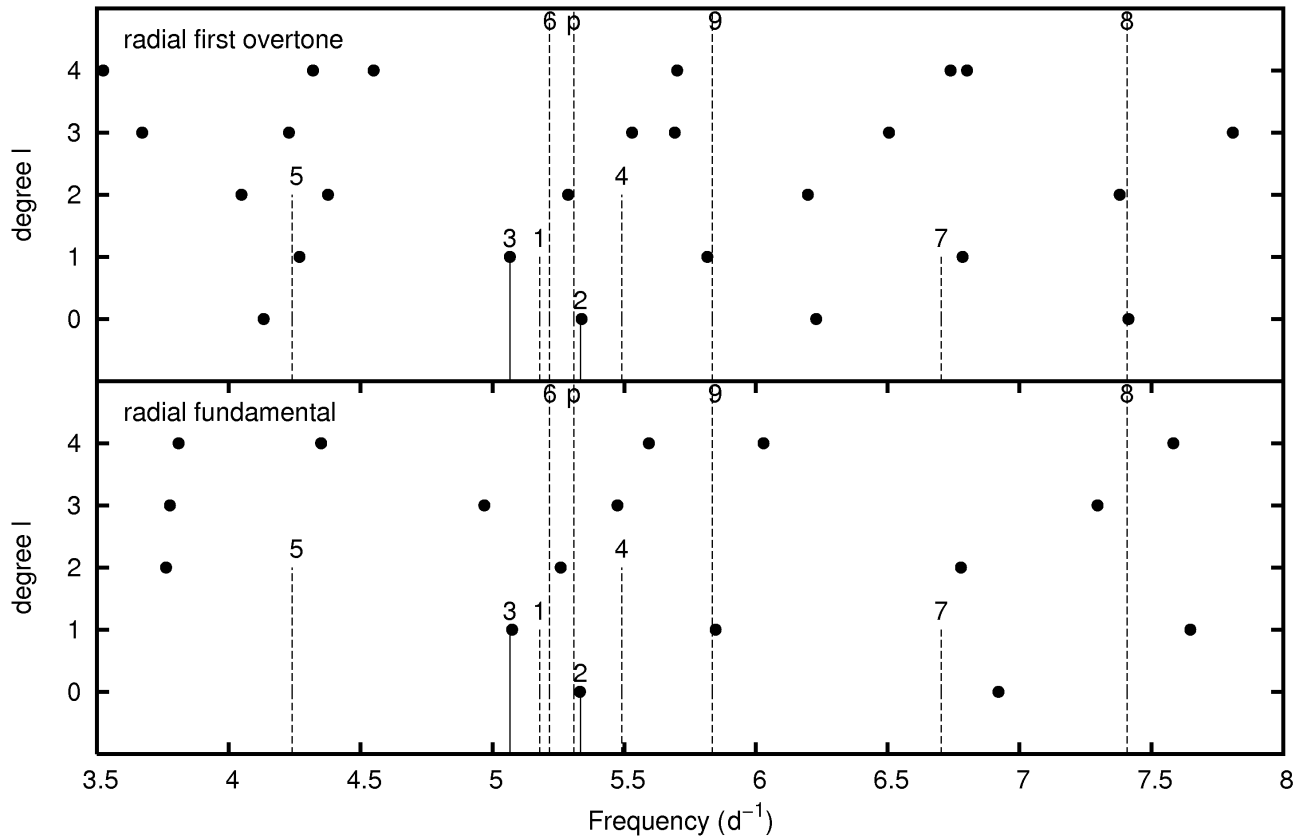}}}
\caption{Comparison between the theoretical frequencies ($m=0$, dots) and the
observed frequencies (full and dotted lines) labeled with their frequency
identifications as listed in Table\,\ref{TAB:LM}. The full lines correspond to the two axisymmetric modes $f_2$ and $f_3$. The top and bottom panels correspond to $f_2$ as the radial first overtone and the radial fundamental, respectively. The theoretical models are calculated for $X = 0.72$, $Z = 0.015$ and $\alpha_{\rm{ov}} = 0.0$.} 
\label{freq_spectrum}
\end{center}
\end{figure*}

As can be seen in Fig.\,\ref{freq_spectrum},
the $(\ell = 1, \rm{p}_3 )$ mode can account for $f_7 = 6.702318$\,\cd.
The match is not perfect for all models in Table\,\ref{model_parameters}. This
is probably due to the fact that there can occur small shifts due to
rotation. On the other hand a 
change in metallicity will also cause a small shift of
the theoretical frequencies (see Sect.\,\ref{sec:disc}).
The frequency $f_8 =
7.407162$\,\cd and $f_9 = 5.84511$\,\cd can only be the $(\ell = 2, \rm{p}_2)$
mode and the $ (\ell = 1, \rm{p}_2 )$ mode, respectively. Indeed, no theoretical
frequency of an $\ell = 1$ zonal mode and an $\ell = 2$ zonal mode is in the
vicinity of $f_8$ and $f_9$, respectively. We also note that an $\l=0$
radial-mode solution for $f_8$ is excluded (see Table\,\ref{TAB:LM}).

The frequency $f_p = 5.30912\,\cd$ found in photometry and the frequency $f_4 =
5.490133$\,\cd were both identified as quadrupole modes and can be accounted for
by the $(\ell = 2, f)$ mode. Indeed, the theoretical Ledoux frequency
splitting 
assuming the averaged internal rotation frequency value $\bar{f}_{\rm rot}$
derived from $f_1 - f_3$ ranges in [0.186-0.190]\,\cd for different overshooting
values, which corresponds to the difference between $f_p$ and $f_4$. The
frequency $f_p$ thus corresponds to $(\ell,m)=(2,0)$ and $f_4$ to
$(\ell,m)=(2,1)$ in accordance with the $m-$identification for $f_4$. Note that
the theoretical frequency value for the $(\ell = 2, f)$ mode is in best
agreement for $\alpha_{\rm{ov}} = 0.0$ and is 5.287\,\cd.

$f_5 = 4.241787$\,\cd being a prograde $\ell = 2$ mode, is uniquely
identified as the $ (\ell = 2, \rm{g}_2 )$ mode. The corresponding theoretical zonal
mode has a frequency between 4.048 and 4.070\,\cd with a Ledoux splitting for
that mode ranging between 0.183 and 0.185\,\cd deduced from 
$\bar{f}_{\rm rot}$ 
and $\beta_{nl}=0.80$.  
In addition, we thus conclude that $m=1$ for this mode.

$f_6 = 5.218075\,\cd$ may be either identified as  $(\ell\,=\,4,\,
\rm{g}_2 )$ 
or $(\ell\,=\,4,\,\rm{g}_1 )$ (see also Fig.\,\ref{freq_spectrum}). 
The models give $\beta_{nl}=0.97$ for this mode.
In the first case, the zonal frequency ranges between
4.38 and 4.55~\cd with a Ledoux splitting of 0.21-0.22\,\cd deduced from
$\bar{f}_{\rm rot}$. This corresponds to an $m$-value of 3 or 4.
In the second case, the zonal frequency ranges between 5.57 and 5.70~\cd
with a Ledoux splitting of 0.20-0.21~\cd. This corresponds to an $m$-value
of -2.

Finally, nothing can be concluded for $f_g = 0.342841\,\cd$ because of the
dense theoretical frequency spectrum in the low-frequency range and because of
the lack of observed constraints on the wavenumbers of this mode. We
additionally note that we cannot definitely exclude this frequency to be
linked with the  
frequency of rotation. For instance,
$f_g/3$ is compatible with our observed surface equatorial
rotation frequency. As explained in
\citet{2001A&A...380..177B,2004A&A...413..273B}, one way to discriminate
between the pulsation and rotation interpretation is to compare the
variability of the considered frequency in lines of different chemical
elements. Unfortunately, we could not achieve such a test because lines other
than Si lines are at our disposal only for two nights of observation.  This
time span is not long enough to recover such a low frequency value.  


\subsection{Comparison with photometry}

We used our best models listed in Table\,\ref{model_parameters} to compare our
results with the photometric mode identification of
\citet{2006MNRAS.365..327H}. We computed theoretical photometric amplitude
ratios for our three main identified pulsation modes ($f_2$, $f_3$, $f_4$;
$f_1$ belongs to $f_3$) following \citet{2003A&A...398..677D}. We computed the
amplitude ratios for Str\"omgren $u$ and $v$, and for Johnson $V$.  For this
computation we must determine the nonadiabatic pulsation mode 
parameters $f_T$ and $f_g$, where $f_T$ corresponds to
the local effective temperature variation and $f_g$
corresponds to the local effective gravity variation, both at
the level of the photosphere. We used the non-adiabatic code {\sc MAD}
\citep{2001A&A...366..166D}
to compute these $f_{T,g}$ parameters for our 
best seismic models of 12~Lac.

\begin{table}
\centering
\caption{Observed photometric amplitude ratios 
taken from \citet{2006MNRAS.365..327H}. We also list our best fitting 
theoretical amplitude ratios. They originate from the model with $\alpha_{\rm
ov}=0.4$ given in Table\,\ref{model_parameters}.}

\label{tab:phot_gerald}
\begin{center}
\begin{tabular}{cccccc}
\hline \hline
ID & Frequency &   \multicolumn{2}{c}{$v/u$} & \multicolumn{2}{c}{$V/u$}  \\
& [\cd] & obs. & theory  &obs. & theory \\
\hline
\ft     & 5.334224 &$  0.529(7) $&0.64 &$ 0.456(7)$& 0.56  \\
\fth    & 5.066316 &$ 0.716(8)  $&0.73 &$0.686(8) $& 0.67  \\
\ff     & 5.490133 &$  0.82(1)  $&0.83 &$0.78(1)  $& 0.78  \\
\hline  
\end{tabular}
\end{center}
\end{table}


We list the observed and theoretical photometric amplitudes for the three main
modes in Table\,\ref{tab:phot_gerald}. We refer to Fig.\,4 of
\citet{2006MNRAS.365..327H} for the results of the photometric mode
identification.  For all the models, we observed that the amplitude ratios
decrease with decreasing mass.  For $f_2$ we find amplitude ratios for $v/u$
ranging from 0.68 until 0.64, and for $V/u$ ranging from 0.51 until 0.56. Thus,
we encounter exactly the same problem as \citet{2006MNRAS.365..327H}, i.e., we
could not find an perfect agreement for the observed ratios of $f_2$. For $f_3$
we find amplitude ratios for $v/u$ ranging from 0.84 until 0.73, and for $V/u$
ranging from 0.74 until 0.67. Our models give amplitude ratios for $f_4$ for
$v/u$ ranging from 0.90 until 0.83, and for $V/u$ ranging from 0.84 until
0.78. Our conclusion is that we can reproduce the observed photometric amplitude
ratios for the two main modes ($f_3$ and $f_4$) from our best seismic
models. Moreover, from this {\it a posteriori\/} consistency check, we favour a
the more evolved seismic models, i.e.,  those with lower mass and
higher core overshoot (see Tables\,\ref{model_parameters} and
\ref{tab:phot_gerald}).

\subsection{Discussion}\label{sec:disc}

We have shown that adopting $n=2$ for the radial mode cannot be excluded as
was done by \citet{2008MNRAS.385.2061D}, and, in fact, explains all observed
frequencies according to their photometric and spectroscopic mode
identification except for $f_g$.  
The identifications of the radial order together with the
additional $\ell$ identifications are summarized in Table \ref{order}. A perfect
agreement in frequency values is not achieved. One reason might be that slightly
different values for $(X,Z)$ than those adopted could lead to a better
agreement.  Another one may be that small frequency shifts due to rotation may
occur as already pointed out by \citet{2008MNRAS.385.2061D}, although the
surface equatorial rotation frequency is only a few percent of the measured
oscillation frequencies.  We found evidence for some degree of differential
rotation from the observed splitting between $f_1$ and $f_3$. 
Using the Ledoux
splitting, all the other
detected frequencies can be well fitted with one and the same 
averaged rotation frequency $\bar{f}_{\rm rot}$
inside the star, where  $\bar{f}_{\rm rot}$ is derived from the model fitting 
of $f_1$ and $f_3$.  In the absence of the observations of multiplets we do not
have enough constraints to derive the internal rotation profile  as defined in 
Equation (\ref{eq:split}) and,
consequently, to determine its effects on the frequencies and frequency
splittings.

\citet{2008MNRAS.385.2061D} partly included rotation effects in their modelling
and proposed a model with the angular rotation velocity more than 4 times higher
in the centre of the star compared to the surface.  
They 
imposed that the radial mode is the fundamental without considering  
alternative identifications. 
Consequently, they could not account for an $\ell =
1$ for $f_7$. As our results are the same than theirs under the assumption of
the radial fundamental mode, we refer to their paper for the properties of the
resulting seismic model in this case without repeating it here 
(see in particular 
their Fig.\,7 and our Fig.\,\ref{freq_spectrum}). 
Their
assumption of non-rigid rotation is based on the fact that $f_5$ cannot be
fitted because the closest $\ell = 2$ mode is too far to be explained by the
splitting due to a uniform rotation. We recall that, on the contrary, our models
reproduce the $\ell = 1$ for $f_7$ and the Ledoux splitting for
$\bar{f}_{\rm rot}$
can explain $f_5$ as
the $(\ell,m,n)=(2,1,-2)$ mode. As additional argument to non-rigid rotation,
\citet{2008MNRAS.385.2061D} stated that their model with uniform rotation has no
identification for $f_6$.  We can explain it as the ($\ell = 4,\rm{g_{2}}$) mode
or the ($\ell = 4,\rm{g_{1}}$) mode in agreement with the photometric
$\ell$-identification.  Finally, even with a non-uniform rotation, they cannot
reproduce $f_5$ appropriately while they used this frequency value to exclude
rigid rotation.

Until now, the only convincing proof of non-rigid rotation has been achieved for
$V836$~Cen \citep{2003Sci...300.1926A,2004A&A...415..251D} and for $\nu$~Eri
\citep{2004MNRAS.350.1022P,2008MNRAS.385.2061D}. The only other star with two
observed multiplets is $\theta$~Oph for which the observed rotational splittings
cannot rule out a rigid rotation model. For 12 Lac, we are able to fit the
observed frequency spectrum using non-rotating stellar models and Ledoux
splittings based on one consistent value $\bar{f}_{\rm rot}$
for the averaged internal rotation
frequency derived from the splitting between $f_1$ and $f_3$. The deviation
between the rotation frequency derived from fitting the splitting $f_1-f_3$ and
from the observed surface rotation frequency points to some degree of
non-rigidity.  With only two components of one and the same multiplet observed,
we are not able to quantify this statement.

Besides our basic stellar modelling, we also checked if non-adiabatic
computations are able to reproduce the excitation of the observed frequencies.
Indeed, recent works revealed shortcomings in the details of the excitation
mechanism for these massive B-type pulsating stars. For instance, for
$\nu$~Eri \citep{2004MNRAS.350.1022P,2004MNRAS.355..352A}, the observed lowest
g-mode and highest p-mode frequencies were not predicted to be excited by
standard models. The same conclusion is found for 12 Lac. Neither the
frequency $f_g$ nor the highest p-mode frequencies ($f_7$ and $f_8$) are
excited by current models, even with OP opacity tables. Since 1991, the
agreement between the non-adiabatic computations for $\beta$~Cephei models and
the observations has improved with subsequent update of opacities \citep[see
e.g.][]{2007MNRAS.375L..21M,Miglio}.  Consequently, a most probable
explanation for the encountered excitation problem is that current opacities
are still underestimated in the region where the driving of pulsation modes
occurs. We refer to \citet{2008MNRAS.385.2061D} for an additional discussion
on this matter.

\section{Conclusions}

Our study was based on 1820 ground-based, high-resolution, high-S/N, multisite
spectroscopic measurements spread over 14 years.  As pointed out in
\citet{2006MNRAS.365..327H} this effort was necessary to identify the
previously unknown azimuthal order of the pulsation modes and thus to be able
to perform a detailed seismic modelling of the star.  We used the \SiIII 4553
\A line to derive the pulsation characteristics of 12 Lac.  In total we find
10 independent frequencies which were also detected in photometry. Worth
mentioning is that we also clearly recover the low-frequency signal.  One of
our aims was to provide a unique identification of as many modes as possible
for 12 Lac. The important result of combining our spectroscopic results with
the ones from the intensive photometric campaign \citep{2006MNRAS.365..327H}
is the unique identification of both \l and \m of the four highest-amplitude
modes. 

With two state-of-the-art methods, the moment method and the FPF method, we
were able to identify the azimuthal order of the four main modes.  We could
also give a constraint on the azimuthal order of the fifth frequency.  Two
main frequencies were identified as axisymmetric modes. One of which has an \l
value of 1 (\fth), the other one is a radial mode (\ft). The two other main
frequencies are identified as $(\l_1,\m_1)=(1,1)$ and $(\l_4,\m_4)=(2,1)$.
The conclusion for the fifth frequency is that the azimuthal order is likely
to be positive.  In addition, the FPF method could also constrain the
inclination, $i=48\pm2\,^{\circ}$, and the surface equatorial rotational
velocity, $\veq=49\pm2\,\kms$.  

The definite identification of four of the observed modes together with some
constraints of the wavenumbers (either $\ell$-values or sign of $m$) on the
other modes allowed us to carry out a detailed seismic modelling, 
under the
assumption of linear oscillation modes. The most
significant outcome of our modelling is the fact that there is a strong
preference that the observed radial mode
(\ft) is the radial first overtone. If \ft is taken as the radial fundamental,
the frequency spectrum of 12 Lac cannot be reproduced satisfactorily.
We do point out that the theory we used, as well as 
\citet{2008MNRAS.385.2061D}, 
does not include nonlinear effects,
while these may be present in 12 Lac \citep[e.g.][]{1992A&A...257..681M}. It
remains to be studied how these would effect mode identification and seismic
modelling, but given the very moderate effect and the limit to only very few
third-order combination frequencies, we do not expect this to alter our
conclusions, just as for the case of $\nu$ Eri, where a linear theory 
was also used \citep[][]{2004MNRAS.350.1022P}.

The complete frequency spectrum of 12 Lac, except the low frequency, can be
fully identified. Our seismic modelling also revealed an excitation problem.
Indeed, the range of frequencies theoretically excited is not large enough
compared to the observations. This might reflect the fact that opacities are
still underestimated in the region where the driving of pulsation modes
occurs. This conclusion is valid for both the OPAL and OP opacities.
  
Finally, our best-fit models indicate that the overshooting parameter
$\alpha_{\rm{ov}}$ has to be lower than 0.4 local pressure scale heights to
get the best agreement with physical parameters derived from spectroscopic
observations. A more refined seismic modelling requires the computation of a
much denser grid of models and will be done in future work.


\section*{Acknowledgments}
MD, MB, WZ and CA acknowledge financial support from the Research Council of
Leuven University under grant GOA/2008/04.  GH has been supported by the
Austrian Fonds zur F\"orderung der wissenschaftlichen Forschung under grants
R12-N02 and P18339-N08.
Part of this work was based on
observations made with the Nordic Optical Telescope, operated on the island of
La Palma jointly by Denmark, Finland, Iceland, Norway, and Sweden, in the
Spanish Observatorio del Roque de los Muchachos of the Instituto de
Astrofisica de Canarias.


\bibliographystyle{mn}
\bibliography{bibMaarten}

\bsp
\label{lastpage}

\end{document}